\documentclass[aps,twocolumn,prc,floatfix,preprintnumbers,superscriptaddress,notitlepage,nofootinbib]{revtex4-1}
\usepackage{dcolumn}
\usepackage{bm}
\usepackage{color}
\usepackage{amssymb}
\usepackage{amsmath}
\usepackage{graphicx}
\usepackage{amsfonts}
\usepackage{slashed}
\usepackage{pstricks}
\usepackage{float}
\counterwithout{table}{section}
\usepackage[colorlinks = true,
linkcolor = blue,
urlcolor  = blue,
citecolor = blue,
anchorcolor = blue]{hyperref}
\usepackage{array}
\allowdisplaybreaks
\usepackage{epsf}

\usepackage{color}
\definecolor{gray}{rgb}{0.6,0.6,0.6}
\definecolor{darkgreen}{rgb}{0.0, 0.545098, 0.0}
\definecolor{darkblue}{rgb}{0.0, 0.0, 0.545098}

\usepackage{color}
\definecolor{BrickRed}{rgb}{0.8, 0.25, 0.33}
\definecolor{gray}{rgb}{0.6,0.6,0.6}
\definecolor{darkgreen}{rgb}{0.0, 0.545098, 0.0}
\definecolor{mypink1}{rgb}{0.858, 0.188, 0.478}

\newcommand{\eps}{{\varepsilon}}
\begin{document}

\title{Three-body universality in the B meson sector}
\author{Yong-Hui~Lin}
\affiliation{Institut für Kernphysik, Technische Universität Darmstadt, 
	64289 Darmstadt, Germany}

\author{Erik~Wilbring}
\affiliation{Helmholtz--Institut f\"ur Strahlen- und Kernphysik (Theorie)\\ 
	and Bethe Center for Theoretical Physics, Universit\"at Bonn, D-53115 Bonn, 
	Germany}

\author{Hai-Long Fu}
\affiliation{CAS Key Laboratory of Theoretical Physics, Institute of Theoretical Physics,\\
Chinese Academy of Sciences, Beijing 100190, China}
\affiliation{School of Physical Sciences, University of Chinese Academy of Sciences, Beijing 100049, China}

\author{Hans-Werner~Hammer}
\affiliation{Institut für Kernphysik, Technische Universität Darmstadt, 
64289 Darmstadt, Germany}
\affiliation{ExtreMe Matter Institute EMMI and Helmholtz Forschungsakademie Hessen f\"ur FAIR (HFHF), GSI Helmholtzzentrum 
für Schwerionenforschung GmbH, 64291 Darmstadt, Germany}

\author{Ulf-G.~Mei{\ss}ner}
\affiliation{Helmholtz--Institut f\"ur Strahlen- und Kernphysik (Theorie)\\ 
   and Bethe Center for Theoretical Physics, Universit\"at Bonn, D-53115 Bonn, Germany}
\affiliation{Institute for Advanced Simulation (IAS-4), 
  Forschungszentrum J\"ulich, D-52425  J\"ulich, Germany}
  \affiliation{Peng Huanwu Collaborative Center for Research and Education, International Institute for Interdisciplinary and Frontiers, Beihang University, Beijing 100191, China}

\date{\today}

\begin{abstract}

The charged exotic mesons $Z_b(10610)$ and $Z'_b(10650)$ 
observed by the Belle collaboration in 2011 are very close to
the $B^* \bar{B}$ and $B^* \bar{B}^*$ thresholds, respectively.
This suggests their interpretation as shallow hadronic molecules
of $B$ and $B^{*}$ mesons. 
Using the masses of the $Z_b(10610)$ and $Z'_b(10650)$
as input, we rule out the possibility for
universal bound states of three $B$ and $B^{*}$ mesons arising from the
Efimov effect based on their spin-isospin structure. 
As a consequence, we can predict the phase shifts for 
the scattering of $B$ and $B^*$ mesons off the exotic mesons
$Z_b(10610)$ and $Z'_b(10650)$  to leading order 
in a non-relativistic effective field theory with contact
interactions based on two-body information alone. 
\end{abstract}

\maketitle

\section{Introduction}
In 2011, the Belle collaboration reported the discovery of two positively charged mesons in the bottomonium
sector, $Z_{b}(10610)$ and $Z'_b(10650)$~\cite{Z_b_discovery_2}. Their existence was subsequently confirmed by two independent Belle measurements~\cite{Belle:2013urd,Belle:2014vzn}. The masses and widths of these states, as listed in the Review of Particle Physics (RPP), are~\cite{ParticleDataGroup:2024cfk}\footnote{Note that in the RPP~\cite{ParticleDataGroup:2024cfk}, the $Z_{b}(10610)$ and $Z'_b(10650)$ are denoted as $T_{b\bar{b}1}(10610)$
and $T_{b\bar{b}1}(10650)^+$, respectively.}
\begin{align}
 M_{Z} = (10607.2 \pm 2.0) \: \mbox{MeV} \:, & \quad \Gamma_{Z} = (18.4 \pm 2.4) \: \mbox{MeV} \:, \nonumber \\
 M_{Z'} = (10652.2 \pm 1.5) \: \mbox{MeV} \:, & \quad \Gamma_{Z'} = (11.5 \pm 2.2) \: \mbox{MeV} \:.
 \label{Belle-values}
\end{align}
From their production and decay channels \cite{Z_b_discovery_2,Belle:2013urd,Belle:2014vzn,Belle:2015upu}, these mesons must be exotic.
Their quark content can not be simply $q \bar{q}$ as for ordinary mesons but must be $b \bar{b} u \bar{d}$.
Soon after their discovery it was proposed that their constituents cluster into
two bottom mesons which are bound due to hadronic forces~\cite{Voloshin_2}. In particular,
both $Z_b$ states were interpreted as hadronic molecules with flavor wave functions
\begin{align}\label{wavefunctions}
 Z_b & = \frac{1}{\sqrt{2}} ( B^* \bar{B} + \bar{B}^* B) \:, \nonumber\\
 Z'_b & = B^* \bar{B}^* \:.
\end{align}
For further analyses in this framework and alternative scenarios 
such as tetraquarks see, e.g., Refs.~\cite{Sun,Yang,Zhang,Cui,Cleven,Cleven_2,Zhao:2021cvg,He:2024aej,Yalikun:2025ssz}
and Refs.~\cite{Cui, Ali, Guo, Bugg, Chen-Liu,Sadl:2021bme,He:2024aej}, respectively.

The molecular interpretation of the $Z_b(10610)$ and $Z'_b(10650)$ is
supported by the fact that their masses are close to the respective open bottom thresholds defined by the flavor
wave functions in Eq.~(\ref{wavefunctions}). Note that the masses
quoted in Eq.~\eqref{Belle-values} are slightly above the
corresponding thresholds. A more sophisticated analysis of the invariant
mass distributions in an effective field theory with bottom meson
loops, however, showed that
the $Z_b$ and  $Z'_b$ poles are below threshold~\cite{Cleven}.
This finding is consistent with a recent analysis of the resonance
signals based on a formalism
consistent with unitarity and analyticity~\cite{Guo:2016bjq,Hanhart:2015cua,Wang:2018jlv,Baru:2019xnh,Krug:2020ufl,Baru:2020ywb}. However, the question of whether the $Z_b$ and $Z'_b$ mesons are virtual states, bound states or resonances has not been answered definitely.
The interplay of $Z_b$ and  $Z'_b$ exchanges with bottom meson loops in
$\Upsilon$ decays was further scrutinized in Refs.~\cite{Chen:2015jgl,Chen:2016mjn}.
For a more detailed discussion of these issues, see
the  review~\cite{Guo:2017jvc}.

Braaten and collaborators argued in Ref.~\cite{Braaten_1} that the
closeness to a bottom meson threshold is necessary but not sufficient for the interpretation as a hadronic
molecule. They used the Born-Oppenheimer approximation
to analyze the substructure of
$Z_b$ and $Z'_b$ and  concluded that for both states the molecule interpretation is viable \cite{Braaten_1}. A recent Born-Oppenheimer study based on the Lattice potential suggests the presence of a near-threshold structure in the $Z$ mass range~~\cite{Prelovsek:2019ywc}. Similarly, a near-threshold signal in the $Z^\prime$ mass range is pointed out in Ref.~\cite{Hoffmann:2024hbz}. 
Arguably, one of
the most detailed investigations of the $Z_b$ states as hadronic molecules was done in
Refs.~\cite{Cleven, Cleven_2} based on an effective field theory with
heavy meson loops that was originally formulated for the charm quark
sector~\cite{Guo:2010ak}.
In this framework, a variety of testable predictions to confirm or rule out the molecular nature of these states
were given. Some of these predictions will be checked at future high-luminosity experiments.

An analysis of the angular distributions showed that the quantum numbers $J^P = 1^+$ are favored for the two $Z_b$
states \cite{Z_b_discovery_2}. In addition, their quark content fixes the isospin to be one. Thus, the quantum numbers
of both $Z_b$ and $Z'_b$ are $I^G (J^{PC}) = 1^+(1^{+-})$ 
\cite{ParticleDataGroup:2024cfk} and the assumption that they are $S$-wave hadronic molecules of two bottom mesons is tenable.
We use an effective field theory with contact interactions to
describe the $Z_b$'s. Since their binding momentum $\gamma = \sqrt{2 \mu B}$ (with binding energy $B$ and reduced mass
of the constituents $\mu$) is much smaller than the pion mass $m_{\pi}$ (or at least of that order in case of the $Z_b$),
the constituent bottom mesons which have masses around $5~\rm{GeV}$ can be treated as non-relativistic point-like particles
which only interact via short-range contact interactions. Thus, one can apply a non-relativistic effective field theory without
explicit pions to this system.
Similar descriptions of two particle $S$-wave molecules in the charm sector can be found in Refs.~\cite{Canham_Hammer_3, Braaten_Kusunoki} concerning the charm meson molecule $X(3872)$ and
in Ref.~\cite{Wilbring_Hammer_Meissner} for the $Z_c(3900)$ whose interpretation as a molecule is still controversial
\cite{Braaten_1}.

This so-called pionless EFT contains only contact interactions and was originally developed for nucleons
which also display shallow bound states such as the deuteron or the triton
\cite{vanKolck:1997ut,van_Kolck_2, Kaplan_3, Kaplan_4}. The expansion parameter
is $Q / m_{\pi}$, where the scale $Q$ is determined by the typical momentum scales 
of the considered process. Depending on the spin-isospin channel, three-body forces may enter already at leading order
in this theory. In the spin-doublet channel of neutron-deuteron scattering, for example, a Wigner-SU(4)-symmetric three-body force
is required at leading order for proper renormalization \cite{Bedaque:1998kg,Bedaque:1998km,Bedaque:1999ve} and the
triton emerges naturally as an Efimov state \cite{Efimov}, while three-body forces are strongly suppressed in
the spin-quartet channel due to the Pauli principle \cite{Bedaque:1997qi,Bedaque:1998mb}.

The Efimov effect describes 
the emergence of shallow three-particle bound states (called \emph{trimers}) in a system with 
resonant interactions characterized by a large scattering length $a$. It can occur if at least two of the three
particle pairs have resonant interactions.
In particular, the Efimov effect occurs in systems of a shallow two-particle bound or virtual state 
of binding momentum $\gamma\sim 1/a$ and a third particle which has resonant
interactions with at least one of the constituents of the dimer.
For $a \rightarrow \infty$, there are infinitely many trimer states with binding energies $B_3^{(n)}$ which are spaced
equidistantly \cite{Efimov}: $B_3^{(n+1)}/B_3^{(n)} =$~const.
The crucial point is that this constant is universal in the sense that it is independent of the details of the short-range
physics in the system. However, its exact value depends on the masses and
spin-isospin quantum numbers of the particles as
well as the number of resonantly interacting pairs.
In a system with finite scattering length, the geometrical spectrum is cut off in the infared and there will
only be a finite number of states but the dependence of the states on the scattering length $a$ is also universal.

Whether or not the Efimov effect plays are role in a three-particle system depends on the particular spin-isospin channel. 
The emergence of the Efimov effect in pionless EFT is closely connected to the requirement of three-body forces for 
renormalization at leading order.\footnote{The case of a covariant formulation was recently investigated
in Ref.~\cite{Epelbaum:2016ffd}, which arrives at some different conclusions concerning the role of three-body forces.} 
The power counting for three-body forces, in turn, can be obtained from an analysis of the
ultraviolet behavior of the corresponding integral equations~\cite{Bedaque:1998km,Bedaque:1999ve,Griesshammer:2005ga}.
The pionless theory contains only contact interactions and is universal. Thus, it
can be applied to all processes with purely short-range interactions
such as low-energy scattering of $D$ and $D^*$ mesons off the $X(3872)$~\cite{Canham_Hammer_3}
or loss processes of ultracold atoms close to a Feshbach resonance~\cite{Hammer}. An overview of the Efimov effect in nuclear and particle physics can be found in Ref.~\cite{Hammer:2010kp}. 

In this work, we assume that the $Z_b$ and  $Z'_b$ are bound states
and predict $\{Z_b , Z'_b\} - \{B , B^*\}$ scattering in the different spin-isospin channels to leading order in $Q$
in pionless EFT. These predictions could, in principle, be tested in the decays of heavier particles into three $B/B^*$ mesons via final state interactions. If the $Z_b$ and  $Z'_b$ are virtual states, this
process does not exist and one needs to look at the more complicated 
three-body scattering of $B/B^*$ mesons.
Moreover, we analyze the different channels with regard to the existence of 
three-body bound states. We note that
bound states of three $\bar{B}/\bar{B}^{*}$ and three $B/B^{*}$ mesons
were previously investigated in 
Refs.~\cite{Garcilazo:2018rwu,Ma:2018vhp,Deng:2024pwm} using quark models and effective field theory methods.
Here, we focus on three-body bound states arising in the 
$\{Z_b , Z'_b\} - \{B , B^*\}$ scattering
channels as a consequence of the Efimov effect~\cite{Efimov}.

The paper is organized as follows: In Sec.~\ref{sec_for}, we write down an extension of pionless EFT for the  $\{Z_b , Z'_b\} - \{B , B^*\}$-system.
The intergral equations for the molecule-meson scattering amplitudes are derived in Sec.~\ref{section_molecule_meson_scattering_amplitudes}
and the relation of the amplitudes to observables is discussed in Sec.~\ref{section_observables}.
Our results and concluding remarks are presented in Secs.~\ref{section_results} and \ref{section_conclusions}, respectively.

\section{Formalism}
\label{sec_for}
To write down an effective Lagrangian density for the  $\{Z_b , Z'_b\} - \{B , B^*\}$-system,
we start by introducing isospin $I = 1/2$ doublets consisting
of the bottom mesons $B$ and $B^*$:
\begin{align}
  B = \begin{pmatrix} B^+ \\ B^0 \end{pmatrix} \:, & \quad \bar{B} = \begin{pmatrix} \bar{B}^0 \\
    B^- \end{pmatrix} \:, \nonumber\\
  B^* = \begin{pmatrix} B^{*+} \\ B^{*0} \end{pmatrix} \:, & \quad \bar{B}^* = \begin{pmatrix} \bar{B}^{*0} \\
    B^{*-} \end{pmatrix} \:,
\end{align}
where the upper components have $I_3 = +1/2$ and the lower ones have $I_3 = -1/2$. Taking into account that both 
$Z_b$ and $Z_b'$ are isospin $1$ states, we write down two isospin-triplets:
\begin{align}
 Z = \begin{pmatrix} Z_1 \\ Z_2 \\ Z_3 \end{pmatrix} \quad \mbox{and} \quad Z' = \begin{pmatrix} Z_1' \\ Z_2' \\ Z_3' \end{pmatrix} \:.
\end{align}
As usual, the physical states, whose electric charges are indicated by the corresponding superscript, are identified as
\begin{align}
 -\frac{1}{\sqrt{2}} \left( Z_1 \: + \: i Z_2 \right) & \equiv Z^+ \quad \mbox{with} \quad I_3 = +1 \:, \nonumber\\
 Z_3 & \equiv Z^0 \quad \mbox{with} \quad I_3 = 0 \:, \nonumber\\
 \frac{1}{\sqrt{2}} \left( Z_1 \: - \: i Z_2 \right) & \equiv Z^- \quad \mbox{with} \quad I_3 = -1 \:,
\end{align}
and analogously for the $Z'$.

Using a similar analysis for spin, we can write down a non-relativistic effective Lagrangian $\mathcal{L}$ up to leading
order (LO). It contains all $B$ and $B^*$ mesons as degrees of freedom. Additionally, there are two auxiliary dimer fields $Z$ and
$Z'$ representing the $Z_b$ and $Z_b'$, respectively. Since we are interested in $Z B$ scattering,\footnote{Note that $ZB$ is used as a placeholder for
  all $\{Z_b,Z'_b\}$ -- $\{B,B^*\}$ scattering processes.}
in general, we have to include
three-body forces as well.
As will be discussed below, however, their explicit form is not required
to leading order.
Taking into account the spin and isospin structure and the
the particle content of the $Z_b$ and $Z'_b$ (cf. Eq.~(\ref{wavefunctions}))
one finds:
\begin{align}\label{Lagrangian}
  &\mathcal{L} ={} B_{\alpha}^{\dagger} \left(i \partial_t + \frac{\nabla^2}{2M_B} \right) B_{\alpha} \: + \: \bar{B}_{\alpha}^{\dagger}
  \left(i \partial_t + \frac{\nabla^2}{2M_B} \right) \bar{B}_{\alpha} \nonumber\\
  &\phantom{xx} + \: B^{* \dagger}_{i \alpha} \left(i \partial_t + \frac{\nabla^2}{2M_{B^*}} \right) B^*_{i \alpha} \: + \: \bar{B}^{* \dagger}_{i \alpha}
  \left(i \partial_t + \frac{\nabla^2}{2M_{B^*}} \right) \bar{B}^*_{i \alpha} \nonumber\\
 &\phantom{xx} + \: Z^{\dagger}_{i A} \Delta Z_{i A} \: + \: Z'^{\dagger}_{i A} \Delta' Z'_{i A} \nonumber\\
  &\phantom{xx} - \: g \Big[ Z^{\dagger}_{i A} \: \bigg(\bar{B}^*_{j \alpha} \: \delta_{ij} (\tau_2 \tau_A)_{\alpha \beta} \: B_{\beta} \: \notag\\
  &\phantom{xxxxx}\quad\quad\quad\quad + \: \bar{B}_{\alpha}
    \: \delta_{ij} (\tau_2 \tau_A)_{\alpha \beta} \: B^*_{j \beta} \bigg) \: + \: h.c. \Big] \nonumber\\
 &\phantom{xx} - \: g' \left[ Z'^{\dagger}_{i A} \: \bar{B}^*_{j \alpha} \: (U_i)_{jk} (\tau_2 \tau_A)_{\alpha \beta} \: B^*_{k \beta} \: + \: h.c. \right] +\ldots \:,
\end{align}
where the ellipsis denotes higher-oder terms,
lowercase Latin letters ($i,j,k... \in \{1,2,3\}$) are spin-$1$ indices,
Greek lowercase letters ($\alpha , \beta , \gamma... \in \{1,2\}$)
are isospin-$1/2$ indices, and uppercase Latin letters
($A,B,C... \in \{1,2,3\}$) denote isospin $1$ for the dimer fields. The matrices $\tau_A$ are Pauli matrices acting in
isospin space and the matrices $U_i$ are the generators of the rotation group acting on the spin-1
representation. Furthermore, we introduce
two coupling constants $g$ and $g'$ for the interaction between the dimer
fields and their constituents. The coefficients of the kinetic terms of $Z$ and $Z'$,
$\Delta$ and $\Delta'$, are also constants.
At leading order in $Q$, $\Delta^{(\prime)}$ and $g^{(\prime)}$ are not independent
and only kept for convenience.
Furthermore, both auxiliary fields are not dynamic. 
However, their bare propagators are dressed by bottom meson loops, so that the full propagators
\begin{align}\label{full_propagators_definition}
 iS_{Z^{(\prime)}} (p_0,\vec{p}) & = \frac{i}{\left(S_{Z^{(\prime)}}^0 \right)^{-1} \: + \: \Sigma^{(\prime)}} \:,
\end{align}
can be expressed in terms of the bare ones $S_{Z,Z'}^0$ and the self-energies $\Sigma$ and $\Sigma'$ which are functions of the
four momentum $p = (p_0, \vec{p})$. They are ultraviolet divergent and need to be regulated using a momentum
cutoff $\Lambda$. Using the reduced masses of $Z_b$ and $Z'_b$,
\begin{align}
 \mu & = \frac{M_B + M_{B^*}}{M_B M_{B^*}} \quad \mbox{and}\quad
 \mu' = \frac{M_{B^*}}{2} \:,
\end{align}
and their kinetic molecule masses $M_Z = M_B + M_{B^*}$ and $M_{Z'} = 2 M_{B^*}$, one can
calculate their self-energies. The self-energy $\Sigma$ of the $Z_b$ is
given by
\begin{align}\label{selfenergy_Z}
  \Sigma (p_0,\vec{p}) & = \frac{2 g^2 \mu}{\pi} \left[- \sqrt{- 2\mu \left(p_0 - \frac{\vec{p}^2}{2M_Z} \right) -i\varepsilon} \:
    + \: \frac{2}{\pi}\Lambda \right] \:,
\end{align}
where $\Lambda$ is a cutoff used to regulate the loop integral for 
the self-energy and $1/\Lambda$ suppressed terms have been neglected.
The self-energy $\Sigma'$ of the $Z'_b$ is obtained from 
Eq.~(\ref{selfenergy_Z}) if all parameters are replaced by their ``primed''
counterparts. Inserting the
self-energies into Eq.~(\ref{full_propagators_definition}) one can match
the scattering amplitudes
\begin{align}
 -iT^{(\prime)} = (-ig^{(\prime)})^2 iS_{Z^{(\prime)}} \left(\frac{k^2}{2\mu^{(\prime)}},0 \right) \:,
\end{align}
with their first order effective range expansions (ERE)
\begin{align}
 \Big(T^{(\prime)}\Big)^{(1)}_{ERE} = -\frac{\pi}{2 \mu^{(\prime)}} \frac{1}{\frac{1}{a^{(\prime)}} + ik} \:,
\end{align}
to obtain the $B$ meson scattering lengths $a$ and $a'$ in the flavor channels of the $Z_b$ and $Z'_b$
(cf.~Eq.~(\ref{wavefunctions})), respectively. We find
\begin{align}
  a^{(\prime)}  = \frac{\pi \Delta^{(\prime)}}{2 \left(g^{(\prime)}\right)^2 \mu^{(\prime)}} \:
  + \: \frac{2}{\pi}\Lambda \:,
\end{align}
where the binding momenta are defined as
\begin{align}\label{binding_momentum_binding_energy_relation}
 \gamma^{(\prime)} &\equiv \frac{1}{a^{(\prime)}}= \mbox{sgn}(B^{(\prime)}) \sqrt{2 \mu^{(\prime)} |B^{(\prime)}|} \:.
\end{align}
Here, the quantity $B=m_1+m_2-M_{12}$ represents the binding energy, which is positive for a bound state and negative for a virtual state. Note, that these definitions are chosen in a way that one takes care of both, bound and virtual states (i.e. a virtual state
corresponds to a negative scattering length). Now one can write the full propagators of both molecules in terms of their binding momentum:
\begin{align}
  &iS_Z (p_0, \vec{p}) = -i \frac{\pi}{2 g^2 \mu} \frac{1}{-\gamma  +  \sqrt{-2\mu \left(p_0  -  \frac{\vec{p}^2}{2 M_Z} \right)  -  i\eps}}\:, \nonumber\\
  &iS_{Z'} (p_0, \vec{p}) = -i \frac{\pi}{2 g'^2 \mu'} \frac{1}{-\gamma'  +  \sqrt{-2\mu' \left(p_0  - 
      \frac{\vec{p}^2}{2 M_{Z'}} \right)  - \: i\eps}} \:. \label{full_propagators_Z}
\end{align}
The wave function renormalization constants, $W$ and $W'$, for both molecules
are given by the residue of the bound state pole of the respective propagators
in Eq.~(\ref{full_propagators_Z}):
\begin{align}
 W^{(\prime)} = \frac{\pi \gamma^{(\prime)}}{2 \left(g^{(\prime)}\right)^2 \left(\mu^{(\prime)}\right)^2} \:.
\end{align}
Higher-order corrections can, in principle, be taken into account by
including additional operators in the Lagrangian
(\ref{Lagrangian})~\cite{vanKolck:1997ut,van_Kolck_2, Kaplan_3, Kaplan_4}. The first correction comes
from the effective range term which is not known for the 
$Z_b$ and $Z'_b$.

\section{Molecule-meson scattering amplitudes}
\label{section_molecule_meson_scattering_amplitudes}
We are interested in the universal properties of the systems of
three $B/B^*$ mesons. This includes scattering processes, such as the
scattering of $B$ and $B^*$ particles off the $Z_b$ and $Z'_b$, as well as bound states
of three $B/B^*$ mesons.
The corresponding information can be extracted from the integral equations for 
$ZB$ scattering where possible bound states appears as
simple poles in the scattering amplitude below threshold.
If such bound states exist, they must be bound due to the Efimov
effect~\cite{Efimov}.

At LO, it is sufficient work with integral equations
for the $ZB$ scattering amplitudes
that contain only two-body interactions.
In channels without shallow trimer states, three-body interactions are
strongly suppressed \cite{Bedaque:1997qi,Bedaque:1998mb}.
Observables become independent of the cutoff $\Lambda$ used to regulate
the loop integrals for large momenta.
If three-body bound states are present and the Efimov effect occurs, however, the
integral equations with two-body interaction only will display a
strong cutoff dependence and a three-body interaction is required for
renormalization already at leading order \cite{Bedaque:1998kg,Bedaque:1998km}.
The running of the three-body interaction is governed by a limit cycle
and thus vanishes at special, log-periodically spaced values of the cutoff.
In particular,
at leading order it is always possible to tune these three-body terms
to zero by working at an appropriate value of the cutoff $\Lambda$~\cite{Hammer:2000nf}.
The value of $\Lambda$ can
then be directly related to the three-body parameter
$\Lambda_*$ which specifies the three-body force~\cite{Bedaque:1998kg,Bedaque:1998km}.
This particular behaviour can also be found using machine learning, see~\cite{Kaspschak:2021hbc}.

The presence of bound states
can therefore be investigated by investigating the cutoff dependence of the
scattering amplitudes in different channels of $ZB$
scattering. If no cutoff dependence is found, shallow three-body
bound states due to the Efimov effect are not present. The $ZB$
scattering amplitudes can then be predicted to leading order
from two-body information alone.

We go on to derive the intergral equations for $ZB$ scattering.
Besides their quark content the isospin doublets $B$ and $\bar{B}$ have the same spin
and isospin degrees of freedom. The small difference in the
masses of their constituents is neglected. If electromagnetic effects are not taken into account, they behave identically
when they are scattered off a $Z$ state. The same argument holds for the doublets $B^*$ and $\bar{B}^*$.
Hence, it is sufficient to analyze the four remaining scattering processes $Z_b B$, $Z_b B^*$, $Z'_b B$
and $Z'_b B^*$. Since both $Z_b$ and $Z'_b$ belong to an isospin-triplet and all relevant bottom mesons
have $I = 1/2$, the isospin structure of all four scattering amplitudes is exactly the same and each
corresponding process has an isospin-$3/2$ and an isospin-$1/2$ channel. In contrast, the spin structure
is different because $B$ and $\bar{B}$ contain pseudoscalar particles while the components of $B^*$ and $\bar{B}^*$
have spin 1. Hence, the $S$-wave scattering of a $B$ off a $Z_b$ or a $Z'_b$ only occurs in a spin-triplet channel
whereas the scattering of a $B^*$ has a spin-singlet, spin-triplet and spin-quintet channel.

%
\begin{figure*}[tb]
\begin{center}
    \includegraphics[width=0.8\textwidth]{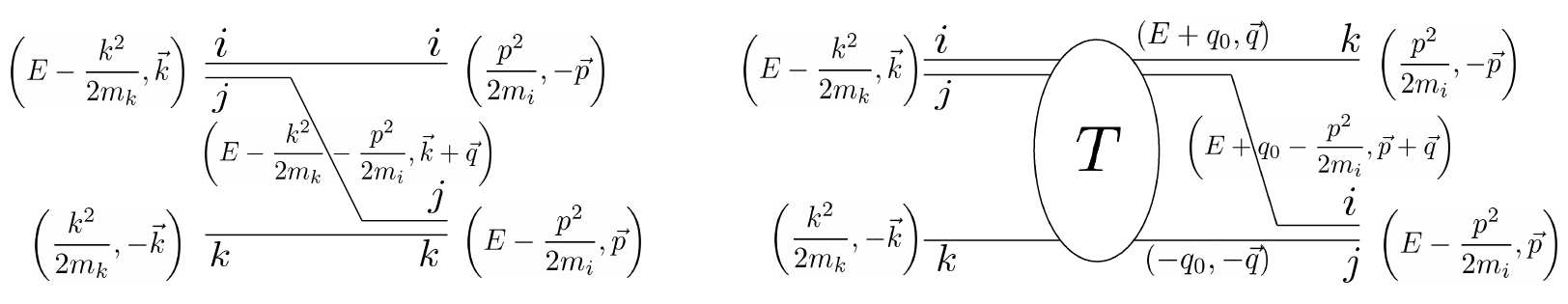}
\end{center}
\caption{\label{Definition_scattering_momenta}{Topologies which appear in the coupled integral equations describing
    molecule-meson scattering. Energies and momenta assigned to the lines are given as
    (energy, momentum). $E$ is the center of mass energy, $\gamma_{ij}$ denotes the binding momentum of the bound state of two mesons $i$ and $j$, and $\mu_{ij}$ is their reduced mass. The associated momenta can be used in
    all $ZB$ scattering processes. }}
\end{figure*}
%

The scattering amplitudes $T$ can be decomposed in a series of partial waves
as
\begin{align}
 T(E,\vec{k},\vec{p}) = \sum_{L = 0}^{\infty} \: (2L + 1) \: T_{(L)} (E,k,p) \: P_{L}(\cos{\theta}) \:,
\end{align}
where $P_{L}$ is a Legendre polynomial,
$\theta$ is the angle between $\vec{k}$ and $\vec{p}$,
$k = |\vec{k}|$, and
$p = |\vec{p}|$. As we focus on trimer states generated by the Efimov effect
which have $L=0$ and meson-molecule scattering processes at low energies, we
project onto $S$-waves and ignore all contributions from higher angular momenta.

It is most convenient to work in the center-of-mass system of the molecule
and the meson.
The coupled integral equations describing molecule-meson scattering contain
only diagrams of the type shown in
Fig.~\ref{Definition_scattering_momenta}. Thus,
we can use the assingment of momenta given in that figure for all $ZB$ channels.
The total energy $E$ is given by
\begin{align}
E = \frac{k^2}{2(m_1 + m_2)} + \frac{k^2}{2m_3} - \frac{\gamma_{12}^2}{2\mu_{12}}
\end{align}
where $\gamma_{ij}$ is the binding momentum of the bound state of two
mesons $i$ and $j$, which has a reduced mass of
$\mu_{ij} = (m_i m_j)/(m_i + m_j)$.
\begin{figure}[tb]
	\begin{center}
		\includegraphics[width=0.48\textwidth]{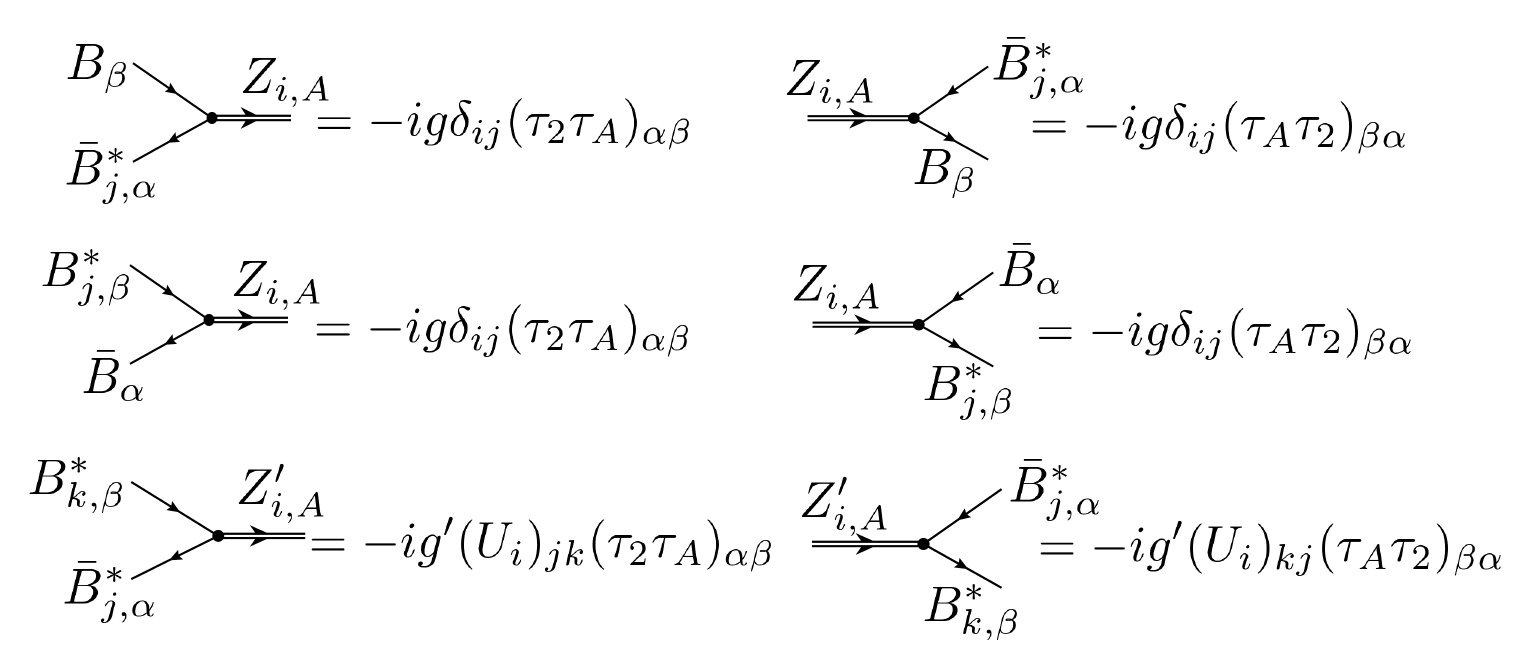}
	\end{center}
	\caption{\label{fig: feynrule}{Feynman rules that follow the Lagrangian density in Eq.~(\ref{Lagrangian}). }}
\end{figure}
%
\subsection{$\mathbf{Z_b B}$ scattering}\label{section_Z_B_scattering}
%
\begin{figure*}[htb]
\begin{center}
    \includegraphics[width=0.7\textwidth]{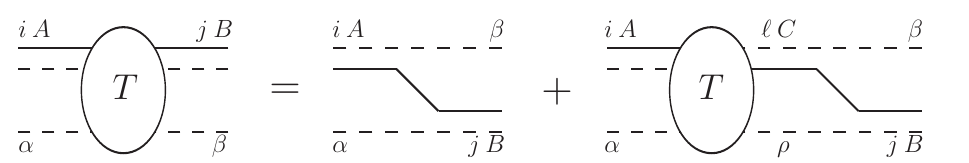}
\end{center}
\caption{\label{Z_B_scattering}{Integral equation of the amplitude $T$ of $Z_b B$ scattering with incoming spin index $i$ and 
    isospin indices $A$, $\alpha$. The respective indices of the outgoing particles are $j$, $B$, and $\beta$. Pseudoscalar mesons
    are depicted as dashed lines and spin-1 bottom mesons as solid lines, respectively.}}
\end{figure*}
%
We start with the simplest scattering process where two of the three bottom mesons are pseudoscalars. The integral equation for the
corresponding scattering amplitude $T$ is shown in Fig.~\ref{Z_B_scattering}. Before proceeding further, let us present the independent tree-level amplitudes for the $Z^{(\prime)}B^{(*)}$ systems. Using the vertex factors, see Fig.~\ref{fig: feynrule}, which follow from the
Lagrangian density Eq.~(\ref{Lagrangian}), we have
\begin{align}\label{eq: tree_amp}
	&i {\cal M}_{ZB\to ZB}^{i A \alpha \to j B \beta} (E, \vec{k}, \vec{p}) ={} \frac{-i g^2\delta_{ij}(\tau_A\tau_B)_{\beta\alpha}}{E-\frac{k^2}{2M_B}-\frac{p^2}{2M_B}-\frac{(\vec{k}+\vec{p})^2}{2M_{B^*}}+i\eps},\notag\\
	&i {\cal M}_{ZB^*\to ZB^*}^{i m A \alpha \to j n B \beta} (E, \vec{k}, \vec{p}) ={} \frac{-i g^2\delta_{j m}\delta_{i n}(\tau_A\tau_B)_{\beta\alpha}}{E-\frac{k^2}{2M_{B^*}}-\frac{p^2}{2M_{B^*}}-\frac{(\vec{k}+\vec{p})^2}{2M_{B}}+i\eps},\notag\\
	&i {\cal M}_{ZB^*\to Z^\prime B}^{i m A \alpha \to j B \beta} (E, \vec{k}, \vec{p}) ={} \frac{-i gg^\prime(U_j)_{im}(\tau_A\tau_B)_{\beta\alpha}}{E-\frac{k^2}{2M_{B^*}}-\frac{p^2}{2M_B}-\frac{(\vec{k}+\vec{p})^2}{2M_{B^*}}+i\eps},\notag\\
	&i {\cal M}_{Z^\prime B\to ZB^*}^{i A \alpha \to j m B \beta} (E, \vec{k}, \vec{p}) ={} \frac{-i gg^\prime(U_i)_{m j}(\tau_A\tau_B)_{\beta\alpha}}{E-\frac{k^2}{2M_B}-\frac{p^2}{2M_{B^*}}-\frac{(\vec{k}+\vec{p})^2}{2M_{B^*}}+i\eps},\notag\\
	&i {\cal M}_{Z^\prime B^*\to Z^\prime B^*}^{i m A \alpha \to j n B \beta} (E, \vec{k}, \vec{p}) ={} \frac{-i g^{\prime 2}(U_i U_j)_{n m}(\tau_A\tau_B)_{\beta\alpha}}{E-\frac{k^2}{2M_{B^*}}-\frac{p^2}{2M_{B^*}}-\frac{(\vec{k}+\vec{p})^2}{2M_{B^*}}+i\eps}.
\end{align}

The integral equation for the $ZB$ scattering is then given by
\begin{align}
  &t_{i A \alpha}^{j B \beta} (E, \vec{k}, \vec{p}) ={} {\cal M}_{ZB\to ZB}^{i A \alpha \to j B \beta} (E, \vec{k}, \vec{p}) \nonumber\\
  &\phantom{xxxxxx} + \: \int \frac{d^4 q}{(2\pi)^4} \:
  \frac{i{\cal M}_{ZB\to ZB}^{\ell C \rho \to j B \beta} (E, \vec{q}, \vec{p})}{-q_0 - \frac{q^2}{2M_B} + i \varepsilon}\notag\\
  &\phantom{xxxxxx}\times \frac{\pi}{2 g^2 \mu}\frac{t_{i A \alpha}^{\ell C \rho} (E, \vec{k}, \vec{q})}{-\gamma + \sqrt{-2\mu \left(E + q_0 - \frac{q^2}{2M_Z} \right)
      - i\varepsilon}},
\end{align}
where $t_{i A \alpha}^{j B \beta}$ is the scattering amplitude including the full
spin-isospin structure.
Integrating over the $q_0$ component and multiplying with the wave function
renormalization, we obtain
\begin{align}
 & T_{i A \alpha}^{j B \beta} (E, \vec{k}, \vec{p}) ={} -\frac{\pi \gamma}{2 \mu^2} \frac{(\tau_A \tau_B)_{\beta\alpha} \:
  	\delta_{ij}}{E - \frac{k^2}{2M_B} - \frac{p^2}{2M_B} - \frac{(\vec{k} + \vec{p})^2}{2M_{B^*}} + i\varepsilon} \nonumber\\
  &\phantom{xxx} - \: \frac{\pi}{2\mu} \: \int \frac{d^3 q}{(2\pi)^3} \:
  \frac{T_{i A \alpha}^{\ell C \rho} (E, \vec{k}, \vec{q})}{-\gamma + \sqrt{-2\mu \left(E - \frac{q^2}{2M_B} - \frac{q^2}{2M_Z} \right)
      - i\varepsilon}} \nonumber\\
 &\phantom{xxx} \times \: \frac{ (\tau_C \tau_B)_{\beta\rho} \: \delta_{\ell j}}{E - \frac{q^2}{2M_B} - \frac{p^2}{2M_B} - \frac{(\vec{q} + \vec{p})^2}{2M_{B^*}} + i\varepsilon} \:,
\end{align}
with $T_{i A \alpha}^{j B \beta} \equiv W \: t_{i A \alpha}^{j B \beta}$.
Evaluating the projection of $T_{i A \alpha}^{j B \beta}$ onto a general
partial wave,
\begin{align}
 \frac{1}{2} \int_{-1}^{1} d\cos \theta \: P_L (\cos \theta) \: T(E, \vec{k}, \vec{p}) = T_{(L)} (E,k,p) \:,
\end{align}
for $L=0$,
we obtain the integral equation for the $S$-wave
$Z_b B$ scattering amplitude
\begin{align}\label{eq: ZB}
  &T_{(0) \, i A \alpha}^{\quad j B \beta} (E,k,p) ={} -\frac{\pi \gamma}{2\mu^2} \boxed{\color{black}(\tau_A \tau_B)_{\beta\alpha} \: \delta_{ij}}\notag\\
 &\phantom{xx}\times \left[-\frac{M_{B^*}}{k p}Q_0\left(-\frac{M_{B^*}}{k p}(E-\frac{k^2}{2\mu}-\frac{p^2}{2\mu})-i\eps \right) \right] \nonumber\\
  &\phantom{} - \frac{1}{4\pi \mu} \int_0^{\Lambda} dq \frac{q^2 \:
    \boxed{\color{black}(\tau_C \tau_B)_{\beta\rho} \: \delta_{\ell j}} \: T_{(0) \, i A \alpha}^{\quad \ell C \rho} (E,k,q)}{-\gamma + \sqrt{-2\mu \left(E - \frac{q^2}{2M_B} - \frac{q^2}{2M_Z} \right)
      - i\varepsilon}} \nonumber\\
  &\phantom{xx}\times \left[-\frac{M_{B^*}}{q p}Q_0\left(-\frac{M_{B^*}}{q p}(E-\frac{q^2}{2\mu}-\frac{p^2}{2\mu})-i\eps \right) \right] \notag\\
  &\equiv \boxed{\color{black}C_{0\;i A \alpha}^{\,\,\, j B \beta}} {\cal M}_0+\int_0^{\Lambda} dq\: {\cal M}_1 \boxed{\color{black}C_{0\; \ell C \rho}^{\,\,\,j B \beta}} T_{(0) \, i A \alpha}^{\quad \ell C \rho} (E,k,q)\:,
\end{align}
where $\Lambda$ is the ultraviolet cutoff discussed above.
The last equality defines the amplitudes ${\cal M}_0$, ${\cal M}_1$ and the coefficients $C_{0\;i A \alpha}^{\,\,\, j B \beta}$ and $C_{0\; \ell C \rho}^{\,\,\,j B \beta}$; for the latter we use boxed notations to make the corresponding definitions more transparent (similar notations will be used later).
The logarithmic function $Q_0$ originates from the one-meson exchange contributions, whose
$S$-wave projection leads to integrals of the type
\begin{align}
 Q_0(\beta)\equiv\frac12\int_{-1}^{+1} dx \frac{P_{0} (x)}{x+\beta} = \frac12\ln \left(\frac{\beta+1}{\beta-1} \right).
\end{align}
\subsubsection{$I = 3/2$, $S = 1$ scattering channel}
We can now choose a specific $Z_bB$ scattering channel. While there is just one spin channel ($S=1$), the isospin $I$ can
either be equal to $3/2$ or equal to $1/2$ since we are coupling
isospin-1 to isospin-$1/2$.
We start with the former. Following Ref.~\cite{Wilbring-Diss},
we project out the desired channel by evaluating:
\begin{align}\label{eq: projection}
	T_{(0)}^{I, S} \equiv \frac{1}{(2S+1)(2I+1)} \sum_{\substack{\tilde{m}\tilde{\eta}, \tilde{n}\tilde{\lambda}}} {\cal O}_{\;\tilde{n}\tilde{\lambda},\tilde{j}\tilde{\beta}}^{\dagger} \: T_{(0) \, \tilde{i}\tilde{\alpha}}^{\quad \tilde{j}\tilde{\beta}} \: {\cal O}_{\tilde{m}\tilde{\eta},\tilde{i}\tilde{\alpha}}\:,
\end{align}
where $\tilde{i}$, $\tilde{j}$, $\tilde{m}$, and $\tilde{n}$ represent general spin indices in the given operators, while $\tilde{\alpha}$, $\tilde{\beta}$, $\tilde{\eta}$, and $\tilde{\lambda}$ denote general isospin indices. Note that for elastic scattering, the initial and final states must be identical, requiring $\tilde{\eta}=\tilde{\lambda}$ and $\tilde{m}=\tilde{n}$.
The projectors for the $Z_b B$ scattering are given by
\begin{align}\label{eq: pw_operators}
	{\cal O}_{j, i}^{S=1}(1\otimes 0\to1)=\:&\delta_{ij}\:,\notag\\
	{\cal O}_{ \beta,A\alpha}^{I=1/2}\left(1\otimes \frac12\to \frac12\right)=\:&\frac{-1}{\sqrt{3}}(\tau_A)_{\alpha\beta}\:,\notag\\
	{\cal O}_{ j\beta,A\alpha}^{I=3/2}\left(1\otimes \frac12\to \frac32\right)=\:&\frac13[(\tau_j\tau_A)_{\alpha\beta}+\delta_{Aj}\delta_{\alpha\beta}]\:.
\end{align}

Applying the above projections
to the $S$-wave $Z_bB$ integral equation of Eq.~(\ref{eq: ZB}), one gets
%
\begin{align}\label{eq: ZB_pw32}
	&T_{(0)}^{I=\frac{3}{2}, S=1} \equiv \frac{1}{12} \sum_{\substack{mD\eta,\\ n E\lambda}} \bigg({\cal O}_{E\lambda, B\beta}^{I=3/2}{\cal O}_{n,j}^{S=1}\bigg)^\dagger  \: T_{(0) \, i A \alpha}^{\quad j B \beta} \notag\\
	&\phantom{xxxxxx}\qquad\qquad\qquad\qquad\times\bigg({\cal O}_{D\eta,A\alpha}^{I=3/2}{\cal O}_{ m,i}^{S=1}\bigg)\notag\\
	&\phantom{}= \frac1{12}\frac19\left[\sum_{m,n}\delta_{jn}\delta_{ij}\delta_{im}\right]\Bigg[\sum_{\substack{D\eta,\\ E\lambda}}\bigg((\tau_B\tau_E)_{\lambda\beta}+\delta_{BE}\delta_{\lambda\beta}\bigg)\notag\\
	&\phantom{xx}\times(\tau_A \tau_B)_{\beta\alpha}\bigg((\tau_D\tau_A)_{\alpha\eta}+\delta_{DA}\delta_{\alpha\eta}\bigg)\Bigg]{\cal M}_0\notag\\
	&\phantom{}+\frac{1}{12}\int_0^{\Lambda} dq\: {\cal M}_1 \sum_{\substack{mD\eta,\\ n E\lambda}} \bigg({\cal O}_{E\lambda, B\beta}^{I=3/2}{\cal O}_{n,j}^{S=1}\bigg)^\dagger  C_{0\; \ell C \rho}^{\,\,\,j B \beta} T_{(0) \, i A \alpha}^{\quad \ell C \rho} \notag\\
	&\phantom{xxxxxx}\qquad\qquad\qquad\qquad\times\bigg({\cal O}_{D\eta,A\alpha}^{I=3/2}{\cal O}_{ m,i}^{S=1}\bigg)\notag\\
	&\phantom{}=\frac1{12}\frac19\times3\times72\: {\cal M}_0+2\int_0^{\Lambda} dq\: {\cal M}_1\frac1{12} \notag\\
	&\phantom{xx}\times\sum_{\substack{mD\eta,\\ n E\lambda}} \bigg({\cal O}_{E\lambda, C\rho}^{I=3/2}{\cal O}_{n,j}^{S=1}\bigg)^\dagger T_{(0) \, i A \alpha}^{\quad j C \rho}\bigg({\cal O}_{D\eta,A\alpha}^{I=3/2}{\cal O}_{ m,i}^{S=1}\bigg)\notag\\
	&\phantom{}=2\:{\cal M}_0+2\int_0^{\Lambda} dq\: {\cal M}_1 T_{(0)}^{I=\frac{3}{2}, S=1},
\end{align}
%
where summation over repeated indices is implied and the identity $\big[(\tau_B\tau_E)_{\lambda\beta}+\delta_{BE}\delta_{\lambda\beta}\big](\tau_C \tau_B)_{\beta\rho}=2\big[(\tau_C\tau_E)_{\lambda\rho}+\delta_{CE}\delta_{\lambda\rho}\big]$ is used in the second-to-last step.
Consequently, the $S$-wave $Z_b B$ scattering
amplitude for $I = 3/2$ and $S = 1$
satisfies the integral equation
\begin{align}\label{eq: ZB_pw32_short}
	&T_{(0)}^{I=\frac{3}{2}, S=1} (E,k,p) ={} -2\times\frac{\pi \gamma}{2\mu^2}\notag\\
	&\phantom{xxxx}\times \left[-\frac{M_{B^*}}{k p}Q_0\left(-\frac{M_{B^*}}{k p}(E-\frac{k^2}{2\mu}-\frac{p^2}{2\mu})-i\eps \right) \right] \nonumber\\
	&\phantom{} - \frac{2}{4\pi \mu} \int_0^{\Lambda} dq \frac{q^2 \:
		T_{(0)}^{I=\frac{3}{2}, S=1} (E,k,q)}{-\gamma + \sqrt{-2\mu \left(E - \frac{q^2}{2M_B} - \frac{q^2}{2M_Z} \right)
			- i\varepsilon}} \nonumber\\
	&\phantom{xxxx}\times \left[-\frac{M_{B^*}}{q p}Q_0\left(-\frac{M_{B^*}}{q p}(E-\frac{q^2}{2\mu}-\frac{p^2}{2\mu})-i\eps \right) \right]\:.
\end{align}

\subsubsection{$I = 1/2$, $S = 1$ scattering channel}
Similarly, one can project onto the second isospin channel
$I = 1/2$ (cf.~Ref.~\cite{Wilbring-Diss}):
%
\begin{align}\label{eq: ZB_pw12}
	&T_{(0)}^{I=\frac{1}{2}, S=1} = \frac1{6}\frac13\left[\sum_{m,n}\delta_{jn}\delta_{ij}\delta_{im}\right]\sum_{\substack{\eta,\lambda}}(\tau_B\tau_A \tau_B\tau_A)_{\lambda\eta}{\cal M}_0\notag\\
	&\phantom{xx}+\frac1{6}\int_0^{\Lambda} dq\: {\cal M}_1 \sum_{\substack{\eta,\lambda}} \bigg(\frac{-1}{\sqrt{3}}(\tau_B\tau_C\tau_B)_{\lambda\rho}\bigg) \delta_{jn}\delta_{j\ell}  T_{(0) \, i A \alpha}^{\quad \ell C \rho} \notag\\
	&\phantom{xxxxxx}\qquad\qquad\qquad\qquad\times\bigg({\cal O}_{\eta,A\alpha}^{I=1/2}{\cal O}_{ m,i}^{S=1}\bigg)\notag\\
	&\phantom{}=\frac1{6}\frac13\times3\times(-6)\: {\cal M}_0+(-1)\int_0^{\Lambda} dq\: {\cal M}_1\frac1{6} \notag\\
	&\phantom{xx}\times\sum_{\substack{m\eta,n\lambda}} \bigg({\cal O}_{\lambda, C\rho}^{I=1/2}{\cal O}_{n,j}^{S=1}\bigg)^\dagger T_{(0) \, i A \alpha}^{\quad j C \rho}\bigg({\cal O}_{\eta,A\alpha}^{I=1/2}{\cal O}_{ m,i}^{S=1}\bigg)\notag\\
	&\phantom{}=-\:{\cal M}_0-\int_0^{\Lambda} dq\: {\cal M}_1 T_{(0)}^{I=\frac{1}{2}, S=1},
\end{align}
%
with the identity $\tau_B\tau_C\tau_B=-\tau_C$. Consequently, the $S$-wave $Z_bB$ scattering amplitude in the $I = 1/2$ and $S = 1$ channel is given by
\begin{align}\label{eq: ZB_pw12_short}
	&T_{(0)}^{I=\frac{1}{2}, S=1} (E,k,p) ={} \frac{\pi \gamma}{2\mu^2}\notag\\
	&\phantom{xxxx}\times \left[-\frac{M_{B^*}}{k p}Q_0\left(-\frac{M_{B^*}}{k p}(E-\frac{k^2}{2\mu}-\frac{p^2}{2\mu})-i\eps \right) \right] \nonumber\\
	&\phantom{xx} + \frac{1}{4\pi \mu} \int_0^{\Lambda} dq \frac{q^2 \:
		T_{(0)}^{I=\frac{1}{2}, S=1} (E,k,q)}{-\gamma + \sqrt{-2\mu \left(E - \frac{q^2}{2M_B} - \frac{q^2}{2M_Z} \right)
			- i\varepsilon}} \nonumber\\
	&\phantom{xxxx}\times \left[-\frac{M_{B^*}}{q p}Q_0\left(-\frac{M_{B^*}}{q p}(E-\frac{q^2}{2\mu}-\frac{p^2}{2\mu})-i\eps \right) \right]\:.
\end{align}
\subsection{$\mathbf{Z_b B^*}$ scattering}\label{section_Z_B_star_scattering}
Next, we consider $Z_b B^*$ scattering.
The pseudoscalar doublet $B$ is then replaced by
the vector doublet $B^*$. Hence,
there are three spin channels: $S=0,1$, and $2$. 
The isospin structure is the same as in $Z_bB$ scattering discussed
in the previous section.
From Fig.~\ref{Z_B_star_scattering}, we find the 
coupled integral equations for the $Z_b B^*$ scattering amplitude $T_1$:
%
\begin{figure*}[tb]
\begin{center}
    \includegraphics[width=0.8\textwidth]{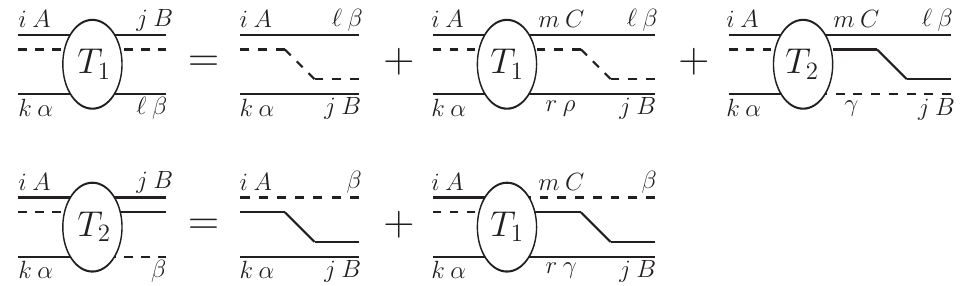}
\end{center}
\caption{\label{Z_B_star_scattering}{Coupled integral equations for the amplitude $T_1$ of $Z_b B^*$ scattering with incoming spin 
    indices $i$, $k$ and isospin indices $A$, $\alpha$, and the corresponding
    outgoing indices
    $j$, $\ell$ and $B$, $\beta$.}}
\end{figure*}
%
%
\begin{align}\label{eq: ZBstar-1}
	&(T_1)_{(0) \, i k A \alpha}^{\quad \, j \ell B \beta} (E,k,p) ={} -\frac{\pi }{2}\frac{\gamma}{\mu^2} \boxed{\color{black}(\tau_A \tau_B)_{\beta\alpha} \: \delta_{jk}\delta_{i\ell}}\notag\\
	&\phantom{xx}\times \left[-\frac{M_{B}}{k p}Q_0\left(-\frac{M_{B}}{k p}(E-\frac{k^2}{2\mu}-\frac{p^2}{2\mu})-i\eps \right) \right] \nonumber\\
	&\phantom{} - \frac{1}{4\pi } \int_0^{\Lambda} dq \frac{q^2 \:
		\left[-\frac{M_{B}}{q p}Q_0\left(-\frac{M_{B}}{q p}(E-\frac{q^2}{2\mu}-\frac{p^2}{2\mu})-i\eps \right) \right]
		}{-\gamma + \sqrt{-2\mu \left(E - \frac{q^2}{2M_{B^*}} - \frac{q^2}{2M_Z} \right)
			- i\varepsilon}} \nonumber\\
	&\phantom{xx}\times \frac{1}{\mu}\boxed{\color{black}(\tau_C \tau_B)_{\beta\rho} \: \delta_{j r}\delta_{\ell m}} \: (T_1)_{(0) \, i k A \alpha}^{\quad m r C \rho} (E,k,q) \notag\\
	&\phantom{} - \frac{1}{4\pi} \int_0^{\Lambda} dq \frac{q^2 \:
		\left[-\frac{M_{B^*}}{q p}Q_0\left(-\frac{M_{B^*}}{q p}(E-\frac{q^2}{2\mu}-\frac{p^2}{2\mu^\prime})-i\eps \right) \right]}{-\gamma^\prime + \sqrt{-2\mu^\prime \left(E - \frac{q^2}{2M_{B}} - \frac{q^2}{2M_{Z^\prime}} \right)
			- i\varepsilon}} \nonumber\\
	&\phantom{xx}\times {\frac{\sqrt{\gamma}}{\mu\sqrt{\gamma^\prime}}}\boxed{\color{black}(\tau_C \tau_B)_{\beta\gamma} \: (U_m)_{\ell j}} \: (T_2)_{(0) \, i k A \alpha}^{\quad m  C \gamma} (E,k,q) \notag\\
	&\equiv \boxed{\color{black}C_{0\;i k A \alpha}^{\,\,\, j \ell B \beta}} {\cal M}_{10}+\int_0^{\Lambda} dq\: {\cal M}_{11} \boxed{\color{black}C_{0\; m r C \rho}^{\,\,\,j \ell B \beta}} (T_1)_{(0) \, i k A \alpha}^{\quad m r C \rho} (E,k,q)\notag\\
	&\phantom{xxx}+\int_0^{\Lambda} dq\: {\cal M}_{12} \boxed{\color{black}C_{1\; m C \gamma}^{\,\,\,j \ell B \beta}} (T_2)_{(0) \, i k A \alpha}^{\quad m C \gamma} (E,k,q)\:,
\end{align}
and
\begin{align}\label{eq: ZBstar-2}
	&(T_2)_{(0) \, i k A \alpha}^{\quad \, j  B \beta} (E,k,p) ={} -\frac{\pi}{2}\frac{\sqrt{\gamma\gamma^\prime}}{\mu\mu^\prime} \boxed{\color{black}(\tau_A \tau_B)_{\beta\alpha} \: (U_j)_{i k}}\notag\\
	&\phantom{xx}\times \left[-\frac{M_{B^*}}{k p}Q_0\left(-\frac{M_{B^*}}{k p}(E-\frac{k^2}{2\mu^\prime}-\frac{p^2}{2\mu})-i\eps \right) \right] \nonumber\\
	&\phantom{} - \frac{1}{4\pi} \int_0^{\Lambda} dq \frac{q^2 \:
		\left[-\frac{M_{B^*}}{q p}Q_0\left(-\frac{M_{B^*}}{q p}(E-\frac{q^2}{2\mu^\prime}-\frac{p^2}{2\mu})-i\eps \right) \right]}{-\gamma + \sqrt{-2\mu \left(E - \frac{q^2}{2M_{B^*}} - \frac{q^2}{2M_{Z}} \right)
			- i\varepsilon}} \nonumber\\
	&\phantom{xx}\times {\frac{\sqrt{\gamma^\prime}}{\mu^\prime\sqrt{\gamma}}}\boxed{\color{black}(\tau_C \tau_B)_{\beta\gamma} \: (U_j)_{m r}} \: (T_1)_{(0) \, i k A \alpha}^{\quad m r C \gamma} (E,k,q) \notag\\
	&\equiv \boxed{\color{black}C_{2\;i k A \alpha}^{\,\,\, j B \beta}} {\cal M}_{20}\notag\\
		&\phantom{xx}+\int_0^{\Lambda} dq\: {\cal M}_{21} \boxed{\color{black}C_{2\; m r C \gamma}^{\,\,\,j B \beta}} (T_1)_{(0) \, i k A \alpha}^{\quad m r C \gamma} (E,k,q)\:,
\end{align}
where the $S$-wave projection and wave function renormalization factors
have already been applied. The projections onto isospin $3/2$
and isospin $1/2$ are the same as in the previous section.
Then the isospin part of the projection operators given by Eq.~\eqref{eq: pw_operators} 
does still work for $Z_b B^*$ scattering.
Since there are three different spin channels which can be combined with both
isospin states, one finds six scattering channels in total. The projectors for the scalar, vector and tensor amplitudes are given by~\cite{Wilbring-Diss}
\begin{align}\label{eq: spin_operator}
	{\cal O}_{ j i}^{S=0}\left(1\otimes 1\to 0\right)=\:&\frac{-1}{\sqrt{3}}\delta_{ij}\:,\notag\\
	{\cal O}_{ \ell,m n}^{S=1}\left(1\otimes 1\to 1\right)=\:&\frac{-1}{\sqrt{2}}(U_\ell)_{m n}\:,\notag\\
	{\cal O}_{\ell k, m n}^{S=2}\left(1\otimes 1\to 2\right)=\:&\frac12[\delta_{\ell m}\delta_{k n}+\delta_{\ell n}\delta_{k m}-\frac23\delta_{\ell k}\delta_{m n}]\:,
\end{align}
where $\left[(U_i)_{jk}\right]^\dagger=(U^\dagger_i)_{kj}=(U_i)_{kj}$ and $(U_i)_{jk}=-i\epsilon_{ijk}$.
Using the same strategy as presented in the $Z_b B$ case, one can obtain the projection coefficients for
the $S$-wave $Z_b B^*$ coupled integral equations by applying the above operators to Eq.~\eqref{eq: ZBstar-1} and \eqref{eq: ZBstar-2}.
The results are collected in Tab.~\ref{Tab: coeff-ZBstar}.
Note that for the inelastic transition $T_2$, the projection is defined as 
\begin{align}
	&(T_2)_{(0)}^{I=1/2,S=1}=\sum_{\substack{g\eta,\\ h \lambda}} \bigg({\cal O}_{\lambda, B\beta}^{I=1/2}{\cal O}_{h,j}^{S=1}\bigg)^\dagger  \: (T_2)_{(0) \, i k A \alpha}^{\quad j B \beta}\notag\\
	&\phantom{xxxxxx}\qquad\qquad\qquad\qquad\times\bigg({\cal O}_{\eta,A\alpha}^{I=1/2}{\cal O}_{g,i k}^{S=1}\bigg)\:.
\end{align}
%
\begin{table}[htbp]
	\centering
	\renewcommand\arraystretch{1.5}
	\caption{Coefficients of the partial-wave projected integral equation for $S$-wave $Z_b B^*$ scattering. One finds
	that $C_0^I=C_1^I=C_2^I$ and $C_1^S$=$C_2^S$, as expected.\label{Tab: coeff-ZBstar}}
	\begin{tabular}{c c c c}
		\hline
		\hline
		Channel & $C_0^IC_0^S$ & $C_1^IC_1^S$ & $C_2^IC_2^S$ \\
		\hline
		$I=1/2$, $S=0$  & $(-1)\times 1$  & $(-1)\times0$	&	$(-1)\times0$ \\
		$I=3/2$, $S=0$	& $2\times 1$		&	$2\times0$		&	$2\times0$		\\
		$I=1/2$, $S=1$  & $(-1)\times(-1)$  & $(-1)\times \sqrt{2}$	&	$(-1)\times\sqrt{2}$ \\
		$I=3/2$, $S=1$  & $2\times(-1)$	 & $2\times\sqrt{2}$		&	$2\times\sqrt{2}$	 \\
		$I=1/2$, $S=2$  & $(-1)\times1$  & $(-1)\times0$	&	$(-1)\times0$ \\
		$I=3/2$, $S=2$  & $2\times1$	 & $2\times0$		&	$2\times0$	 \\
		\hline
		\hline
	\end{tabular}
\end{table}
As expected, there is no $T_2$ contribution to the $S=0$ and $S=2$ channel because the scattered particles in this channel
can only couple to a total spin of $S=1$, i.e., $C_1^{S=0,2}=C_2^{S=0,2}=0$. The final integral equations for all six scattering
channels are then expressed in the following general form with the spin-isospin factors listed in Tab.~\ref{Tab: coeff-ZBstar},
\begin{align}\label{eq: ZBstar_pw}
	&(T_1)_{(0)}^{I, S}(E,k,p) = C_0^IC_0^S\:{\cal M}_{10}(k,p)\notag\\
	&\phantom{xxx}+\int_0^{\Lambda} dq\: C_0^IC_0^S {\cal M}_{11}(p,q) (T_1)_{(0)}^{I, S}(E,k,q)\notag\\
	&\phantom{xxx}+\int_0^{\Lambda} dq\: C_1^IC_1^S {\cal M}_{12}(p,q) (T_2)_{(0)}^{I, S}(E,k,q)\:,\notag\\
	&(T_2)_{(0)}^{I, S}(E,k,p) = C_2^IC_2^S\:{\cal M}_{20}(k,p)\notag\\
	&\phantom{xxx}+\int_0^{\Lambda} dq\: C_2^IC_2^S {\cal M}_{21}(p,q) (T_1)_{(0)}^{I, S}(E,k,q)\:.
\end{align}
The five scalar amplitudes ${\cal M}_{10}$, ${\cal M}_{11}$, ${\cal M}_{12}$, ${\cal M}_{20}$ and ${\cal M}_{21}$ are defined by
Eq.~\eqref{eq: ZBstar-1} and \eqref{eq: ZBstar-2}.

\subsection{$\mathbf{Z'_b B}$ scattering}
%
\begin{figure*}[tb]
\begin{center}
    \includegraphics[width=0.8\textwidth]{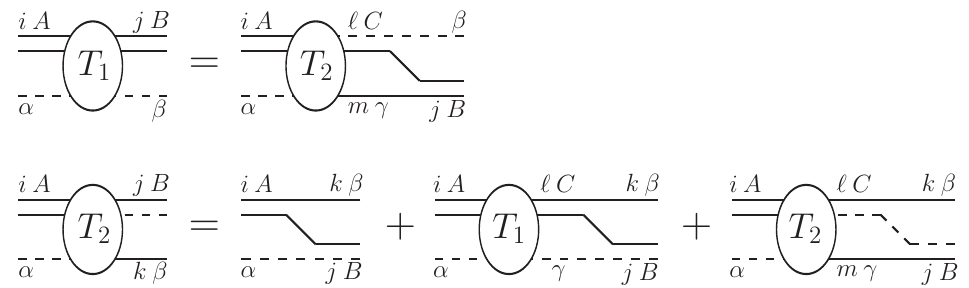}
\end{center}
\caption{\label{Z_prime_B_scattering}{Coupled integral equations for the amplitude $T_1$ of $Z'_b B$ scattering
    with incoming spin index $i$ and isospin indices $A$, $\alpha$,
    and the corresponding outgoing indices
    and $j$ and  $B$, $\beta$.}}
\end{figure*}
%
The scattering process $Z'_bB \rightarrow Z'_bB$ has the same isospin structure as described in the previous sections. This
yields the isospin factors $C_0^I$ in the two amplitudes shown in
Fig.~\ref{Z_prime_B_scattering}.
Although the only spin channel is $S=1$ and thus the projection operator is the same as in
Sec.~\ref{section_Z_B_scattering},
the vertices appearing in the amplitudes are different. This leads to different spin
factors in these amplitudes. Firstly, the $S$-wave $Z'_b B$ scattering amplitudes satisfy the integral equation,
\begin{align}\label{eq: ZpB-1}
	&(T_1)_{(0) \, i A \alpha}^{\quad \, j B \beta} (E,k,p) ={} - \frac{1}{4\pi } \int_0^{\Lambda} dq \notag\\
	&\phantom{xx}\times\frac{q^2 \:
		\left[-\frac{M_{B^*}}{q p}Q_0\left(-\frac{M_{B^*}}{q p}(E-\frac{q^2}{2\mu^\prime}-\frac{p^2}{2\mu})-i\eps \right) \right]
	}{-\gamma + \sqrt{-2\mu \left(E - \frac{q^2}{2M_{B^*}} - \frac{q^2}{2M_Z} \right)
			- i\varepsilon}} \nonumber\\
	&\phantom{xx}\times \frac{\sqrt{\gamma^\prime}}{\mu^\prime\sqrt{\gamma}}\boxed{\color{black}(\tau_C \tau_B)_{\beta\gamma} \: (U_j)_{\ell m}} \: (T_2)_{(0) \, i A \alpha}^{\quad \ell m C \gamma} (E,k,q) \notag\\
	&\equiv +\int_0^{\Lambda} dq\: {\cal M}_{12}\: \boxed{\color{black}C_{3\; \ell m C \gamma}^{\,\,\, j B \beta}} (T_2)_{(0) \, i A \alpha}^{\quad \ell m C \gamma} (E,k,q)\:,
\end{align}
and
\begin{align}\label{eq: ZpB-2}
	&(T_2)_{(0) \, i A \alpha}^{\quad \, j k B \beta} (E,k,p) ={} -\frac{\pi}{2}\frac{\sqrt{\gamma\gamma^\prime}}{\mu\mu^\prime} \boxed{\color{black}(\tau_A \tau_B)_{\beta\alpha} \: (U_i)_{k j}}\notag\\
	&\phantom{xx}\times \left[-\frac{M_{B^*}}{k p}Q_0\left(-\frac{M_{B^*}}{k p}(E-\frac{k^2}{2\mu}-\frac{p^2}{2\mu^\prime})-i\eps \right) \right] \nonumber\\
	&\phantom{} - \frac{1}{4\pi} \int_0^{\Lambda} dq \frac{q^2 \:
		\left[-\frac{M_{B^*}}{q p}Q_0\left(-\frac{M_{B^*}}{q p}(E-\frac{q^2}{2\mu}-\frac{p^2}{2\mu^\prime})-i\eps \right) \right]}{-\gamma^\prime + \sqrt{-2\mu^\prime \left(E - \frac{q^2}{2M_{B}} - \frac{q^2}{2M_{Z^\prime}} \right)
			- i\varepsilon}} \nonumber\\
	&\phantom{xx}\times {\frac{\sqrt{\gamma}}{\mu\sqrt{\gamma^\prime}}}\boxed{\color{black}(\tau_C \tau_B)_{\beta\gamma} \: (U_\ell)_{k j}} \: (T_1)_{(0) \, i A \alpha}^{\quad \ell C \gamma} (E,k,q) \notag\\
	&\phantom{} - \frac{1}{4\pi} \int_0^{\Lambda} dq \frac{q^2 \:
		\left[-\frac{M_{B}}{q p}Q_0\left(-\frac{M_{B}}{q p}(E-\frac{q^2}{2\mu}-\frac{p^2}{2\mu})-i\eps \right) \right]}{-\gamma + \sqrt{-2\mu \left(E - \frac{q^2}{2M_{B^*}} - \frac{q^2}{2M_{Z}} \right)
			- i\varepsilon}} \nonumber\\
	&\phantom{xx}\times {\frac{1}{\mu}}\boxed{\color{black}(\tau_C \tau_B)_{\beta\gamma} \: \delta_{j m}\delta_{\ell k}} \: (T_2)_{(0) \, i A \alpha}^{\quad \ell m C \gamma} (E,k,q) \notag\\
	&\equiv \boxed{\color{black}C_{4\;i A \alpha}^{\,\,\, j k B \beta}} {\cal M}_{20}\notag\\
	&\phantom{xx}+\int_0^{\Lambda} dq\: {\cal M}_{21} \boxed{\color{black}C_{4\; \ell C \gamma}^{\,\,\,jk B \beta}} (T_1)_{(0) \, i A \alpha}^{\quad \ell C \gamma} (E,k,q)\notag\\
	&\phantom{xx}+\int_0^{\Lambda} dq\: {\cal M}_{22} \boxed{\color{black}C_{5\; \ell m C \gamma}^{\,\,\,jk B \beta}} (T_2)_{(0) \, i A \alpha}^{\quad \ell m C \gamma} (E,k,q)\:,
\end{align}
Similarly, by applying the spin-isospin projection onto these equations, we obtain
\begin{align}\label{eq: ZpB_pw}
	&(T_1)_{(0)}^{I, S}(E,k,p) = +\int_0^{\Lambda} dq\: C_3^IC_3^S {\cal M}_{12}(p,q) (T_2)_{(0)}^{I, S}(E,k,q),\notag\\
	&(T_2)_{(0)}^{I, S}(E,k,p) = C_4^IC_4^S\:{\cal M}_{20}(k,p)\notag\\
	&\phantom{xxx}+\int_0^{\Lambda} dq\: C_4^IC_4^S {\cal M}_{21}(p,q) (T_1)_{(0)}^{I, S}(E,k,q)\notag\\
	&\phantom{xxx}+\int_0^{\Lambda} dq\: C_5^IC_5^S {\cal M}_{22}(p,q) (T_2)_{(0)}^{I, S}(E,k,q)\:,
\end{align}
with four scalar amplitudes ${\cal M}_{12}$, ${\cal M}_{20}$, ${\cal M}_{21}$ and ${\cal M}_{22}$ whose expressions are
given by Eq.~\eqref{eq: ZpB-1} and \eqref{eq: ZpB-2}. Note that $C_3^I=C_4^I=C_5^I=C_0^I$. The spin coefficients are given by
Tab.~\ref{Tab: coeff-ZpB}.
\begin{table}[htbp]
	\centering
	\renewcommand\arraystretch{1.5}
	\caption{Coefficients of the partial-wave projected integral equation for $S$-wave $Z'_b B$ scattering.\label{Tab: coeff-ZpB}}
	\begin{tabular}{c c c c}
		\hline
		\hline
		Channel & $C_3^IC_3^S$ & $C_4^IC_4^S$ & $C_5^IC_5^S$ \\
		\hline
		$I=1/2$, $S=1$  & $(-1)\times\sqrt{2}$  & $(-1)\times \sqrt{2}$	&	$(-1)\times(-1)$ \\
		$I=3/2$, $S=1$  & $2\times\sqrt{2}$	 & $2\times\sqrt{2}$		&	$2\times(-1)$	 \\
		\hline
		\hline
	\end{tabular}
\end{table}
%
\subsection{$\mathbf{Z'_b B^*}$ scattering}
%
\begin{figure*}[tb]
\begin{center}
    \includegraphics[width=0.7\textwidth]{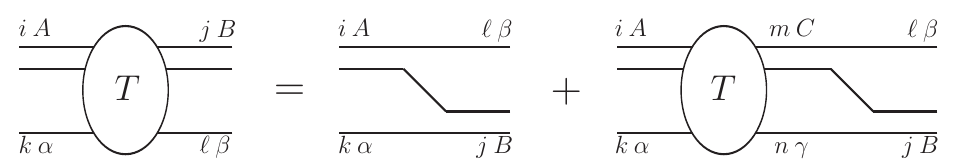}
\end{center}
\caption{\label{Z_prime_B_star_scattering}{Integral equation for the amplitude $T$ of $Z'_b B^*$ scattering with
    incoming spin indices $i$, $k$ and isospin indices $A$, $\alpha$,
    and the corresponding outgoing indices
    $j$, $\ell$ and $B$, $\beta$.}}
\end{figure*}
%
The last scattering process includes two vector particles. Thus, there are six channels in total (three spin states $S=0,1,2$
combined with the isospin $1/2$ and $3/2$ states). The integral equation for this process is shown in
Fig.~\ref{Z_prime_B_star_scattering}. After wave function renormalization and $S$-wave projection we find
\begin{align}\label{eq: ZpBstar}
	&T_{(0) \, ik A \alpha}^{\quad \, j \ell B \beta} (E,k,p) ={} -\frac{\pi}{2}\frac{\gamma^\prime}{\mu^{\prime\:2}} \boxed{\color{black}(\tau_A \tau_B)_{\beta\alpha} \: (U_iU_j)_{\ell k}}\notag\\
	&\phantom{xx}\times \left[-\frac{M_{B^*}}{k p}Q_0\left(-\frac{M_{B^*}}{k p}(E-\frac{k^2}{2\mu^\prime}-\frac{p^2}{2\mu^\prime})-i\eps \right) \right] \nonumber\\
	&\phantom{} - \frac{1}{4\pi} \int_0^{\Lambda} dq \frac{q^2 \:
		\left[-\frac{M_{B^*}}{q p}Q_0\left(-\frac{M_{B^*}}{q p}(E-\frac{q^2}{2\mu^\prime}-\frac{p^2}{2\mu^\prime})-i\eps \right) \right]}{-\gamma^\prime + \sqrt{-2\mu^\prime \left(E - \frac{q^2}{2M_{B^*}} - \frac{q^2}{2M_{Z^\prime}} \right)
			- i\varepsilon}} \nonumber\\
	&\phantom{xx}\times {\frac{1}{\mu^\prime}}\boxed{\color{black}(\tau_C \tau_B)_{\beta\gamma} \: (U_mU_j)_{\ell n}} \: T_{(0) \, i k A \alpha}^{\quad mn C \gamma} (E,k,q) \notag\\
	&\equiv \boxed{\color{black}C_{6\;ik A \alpha}^{\,\,\, j \ell B \beta}} {\cal M}_{0}+\int_0^{\Lambda} dq\: {\cal M}_{1} \boxed{\color{black}C_{6\; m n C \gamma}^{\,\,\,j\ell B \beta}} T_{(0) \, i k A \alpha}^{\quad mn C \gamma} (E,k,q)\:,
\end{align}
The spin projection operators are known from Sec.~\ref{section_Z_B_star_scattering} and given in Eq.~\eqref{eq: spin_operator}. When applied to the amplitude
of the above equation one ends up with the following spin-isospin factors, see Tab.~\ref{Tab: coeff-ZpBstar}.
\begin{table}[htbp]
	\centering
	\renewcommand\arraystretch{1.5}
	\caption{Coefficients of the partial-wave projected integral equation for $S$-wave $Z'_b B^*$ scattering.\label{Tab: coeff-ZpBstar}}
	\begin{tabular}{c c c c}
		\hline
		\hline
		$(I,S)$ 		& $(\frac12,0)$ 		& $(\frac12,1)$		& $(\frac12,2)$  \\
		$C_6^I C_6^S$ 	& $(-1)\times (-2)$   	& $(-1)\times 1$   	& $(-1)\times1$\\
		\hline
		$(I,S)$ 		& $(\frac32,0)$ 		& $(\frac32,1)$ 	&$(\frac32,2)$ \\
		$C_6^I C_6^S$ 	& $2\times (-2)$ 		&$2\times 1$  		& $2\times1$ \\
		\hline
		\hline
	\end{tabular}
\end{table}
Then the projected integral equation for the $S$-wave $Z'_b B^*$ scattering reads
\begin{align}\label{eq: ZpBstar_pw}
	&T_{(0)}^{I, S}(E,k,p) = C_6^IC_6^S\:{\cal M}_{0}(k,p)\notag\\
	&\phantom{xxxxxx}+\int_0^{\Lambda} dq\: C_6^IC_6^S {\cal M}_{1}(p,q) T_{(0)}^{I, S}(E,k,q)\:,
\end{align}
with two scalar amplitudes ${\cal M}_{0}$ and ${\cal M}_{1}$ defined by Eq.~\eqref{eq: ZpBstar}.

\section{Relation to Observables}
\label{section_observables}
After discretization, the integral equations
from Sec.~\ref{section_molecule_meson_scattering_amplitudes}
reduce to inhomogeneous matrix equations of the form
$\mathbf{T} = \mathbf{R} \: + \: \mathcal{M} \mathbf{T}$,
where all quantities implicitly depend on the energy $E$. For an integral equation describing
$ZB$ scattering in a single channel there are three relevant regions of the center of mass energy $E$. Denoting the binding energy of
the relevant molecule $Z_b$ or $Z'_b$ as $B(Z)$, we have:
\begin{enumerate}
\item[(i)] $-B(Z) \leq E \leq 0$: here,
  the elastic scattering of a bottom meson off a molecule is the only process that is allowed.
  In terms of the center-of-mass momentum $k$ this energy region translates to $0 \leq k \leq k_{break}$,
  where $k_{break}$ is the breakup momentum of the molecule.
\item[(ii)] $-\infty < E < -B(Z)$: for energies below the molecule-meson scattering threshold, trimer states can appear.
\item[(iii)] $0 < E < \infty$: for positive energies the molecule can break apart and three-particle singularities have to be 
  taken into account. This regime is beyond the scope of our work.
\end{enumerate}
In a system of two coupled integral equations where both $Z_b$ and $Z'_b$ are involved, the scenario above must
be generalized. One has to replace in the second
case $-\infty < E < -B(Z)$ by $-\infty < E < -\max (B,B')$ because a stable trimer state must lie below both dimer thresholds.
In the first case one has to take care of the relation between the two binding energies $B$ and $B'$. Purely elastic two-body
scattering $Z_b B^*$, for example, only takes place for $B \geq B'$.
Namely, in the energy region $-B \leq E \leq -B'$. In the other case, i.e. for $B < B'$ both molecule states can be formed
out of the three bottom mesons and inelastic reactions become possible. Such reactions will not be considered here.
In $Z'_b B$ scattering, the situation is reversed.

In the following, we consider a system of two coupled integral equations of the type derived in
Sec.~\ref{section_molecule_meson_scattering_amplitudes}. The simpler case with just one such equation can
straightforwardly be deduced from this.
\subsection{Elastic $\mathbf{ZB}$ scattering}
\label{section_elastic_ZB_scattering}
In the energy region (i), where elastic $ZB$ scattering is dominant,
the corresponding amplitudes $T_1 (p)$ and $T_2 (p)$
can be found by solving the inhomogeneous matrix equation $\mathbf{T} = \mathbf{R} \: + \: \mathcal{M} \mathbf{T}$ for
a given momentum $k$, i.e. at a given center of mass energy $E \sim k^2$.
Besides the amplitudes themselves there are two additional
observables of interest in the $ZB$ scattering process: the meson-molecule scattering length $a_3$ and the phase shift $\delta_{L}(k)$
where $L$ is the relative angular momentum between  $Z$ and $B$.
Since we focus on $S$-wave scattering, we define the $S$-wave
phase shift as $\delta (k) \equiv \delta_0 (k)$.

For the determination of these two quantities, we use the relation
\begin{align}\label{general_scattering_amplitude}
 T_1(k,p=k) = \frac{2\pi}{\mu_3} \frac{1}{k \cot \delta - ik} \:,
\end{align}
with the effective range expansion
\begin{align}
 k \cot \delta = -\frac{1}{a_3} + \mathcal{O}(k^2) \:. 
\end{align}
Thus, the scattering length $a_3$ is given by
\begin{align}\label{three_body_scattering_length_def}
 a_3 = -\frac{\mu_3}{2\pi} \: T_1(0,0) \:,
\end{align}
and the scattering phase shift can be determined by inverting Eq.~(\ref{general_scattering_amplitude}).
\subsection{Trimer states}
\label{section_three_particle_bound_states}
For negative energies below the two particle threshold, there are no poles in the kernels of the integral equations.
A three-particle bound state with binding energy $B_3$ shows up as a simple pole in the two amplitudes $T_1$ and $T_2$ which are combined in $\mathbf{T}$. One can parametrize the amplitudes in the vicinity of the pole as
\begin{align}
 T_1 (k,p) & = \frac{B(k) \: B_1(p)}{E + B_3} + \mbox{regular terms,}\nonumber \\
 T_2 (k,p) & = \frac{B(k) \: B_2(p)}{E + B_3} +\mbox{regular terms,} \quad \mbox{for $E \rightarrow -B_3$} \:.
\end{align}
Inserting this into the coupled integral equations for $T_1 (k,p)$ and  $T_2 (k,p)$ and
matching the coefficients of the pole in $(E+B_3)$, we
 obtain a homogeneous integral equation for $B(p)$
 which has nontrivial solutions only for a discrete (and possibly empty) set of negative bound state energies.
 After discretization, this turns into a homogeneous matrix equation of the
form $\mathbf{B} = \mathcal{M}(E) \mathbf{B}$.
\section{Results}
\label{section_results}
The question of whether the $Z_b$ and $Z'_b$ mesons are virtual states, bound states or resonances has not been answered definitely
(see, e.g., Ref.~\cite{Baru:2019xnh} for an analysis of recent experimental data on the production and decay channels of the $Z$ and $Z^\prime$ in an effective field theory framework that incorporates constraints from unitarity and analyticity). Here we assume that the $Z$ and $Z^\prime$
are bound states and solve the (coupled) integral equations derived in the previous section. Since their binding energies, required as input for 
these calculations, are uncertain, we follow  the strategy of Ref.~\cite{Baru:2017gwo}
and assume the ranges:
\begin{align}
 B &= 5.0\pm2.5 \: \mbox{MeV} \:, \nonumber \\
 B' &= 1.0\pm0.5 \: \mbox{MeV} \:.
 \label{eq:binding_energies}
\end{align}
Using Eq.~(\ref{binding_momentum_binding_energy_relation}), this leads to the binding momenta
\begin{align}
 \gamma &= 162.8^{+36.6}_{-47.7} \: \mbox{MeV} \:, \nonumber \\
 \gamma' &= 73.0^{+16.4}_{-21.4} \: \mbox{MeV} \:.
\end{align}
for the $Z_b$ and $Z'_b$ molecules, respectively.
One observes that the central value for the $Z_b(10610)$ is larger than the pion mass, so the applicability of an EFT without explicit pions is not obviuous. However, due to the large uncertainties of $\gamma$, a binding momentum of the $Z_b$ below $M_{\pi}$ is not excluded. As a consequence,
one can use pionless EFT as a model to obtain first insights on the properties of the $ZB$ systems. The sensitivity to the input values is illustrated by showing
results for the central values and the upper and lower bounds in Eq.~(\ref{eq:binding_energies}).
As discussed in detail in Ref.~\cite{Guo:2017jvc}, there is some
uncertainty about the precise location of these poles, so our analysis should be updated when precise data becomes available.
\subsection{Bound states of three $\mathbf{B/B^*}$ mesons}
We searched for solutions of the homogeneous integral equations 
(cf.~Subsec.~\ref{section_three_particle_bound_states})
corresponding to bound states of three $B/B^*$ mesons in all
spin and isospin channels of the $Z_b B$ and 
$Z'_b B\,(I=1/2,3/2,\,S=1)$, as well as
$Z_b B^*$ and $Z'_b B^*\,(I=1/2,3/2,\,S=0,1,2)$ systems discussed in Sec.~\ref{section_molecule_meson_scattering_amplitudes}, repectively. No such solutions were found.
As a consequence, there is no Efimov effect with three $B/B^*$ mesons.
Heuristically, this can be understood from the effective number of interacting
pairs, which is smaller than two in all channels. Moreover,
the amplitudes are independent of the cutoff $\Lambda$ for
sufficiently large $\Lambda$ and three-body forces do not enter
at leading order.

Next we focus on $ZB$ scattering in the different
channels.
Due to the suppression of three-body forces, this
is completely predicted by the $Z_b$ and $Z'_b$ binding
energies to leading order and the cutoff $\Lambda$ in the integral equations
in Sec.~\ref{section_molecule_meson_scattering_amplitudes} can be
removed. Note that we will not show numerical results for
$Z'_b B$ scattering, since 
a purely
elastic scattering process without coupling to the $Z_b B^*$ 
system is not possible
(cf.~the discussion in Sec.~\ref{section_observables}).
%
%
\begin{figure}[tb]
\begin{center}
    \includegraphics[width=0.48\textwidth]{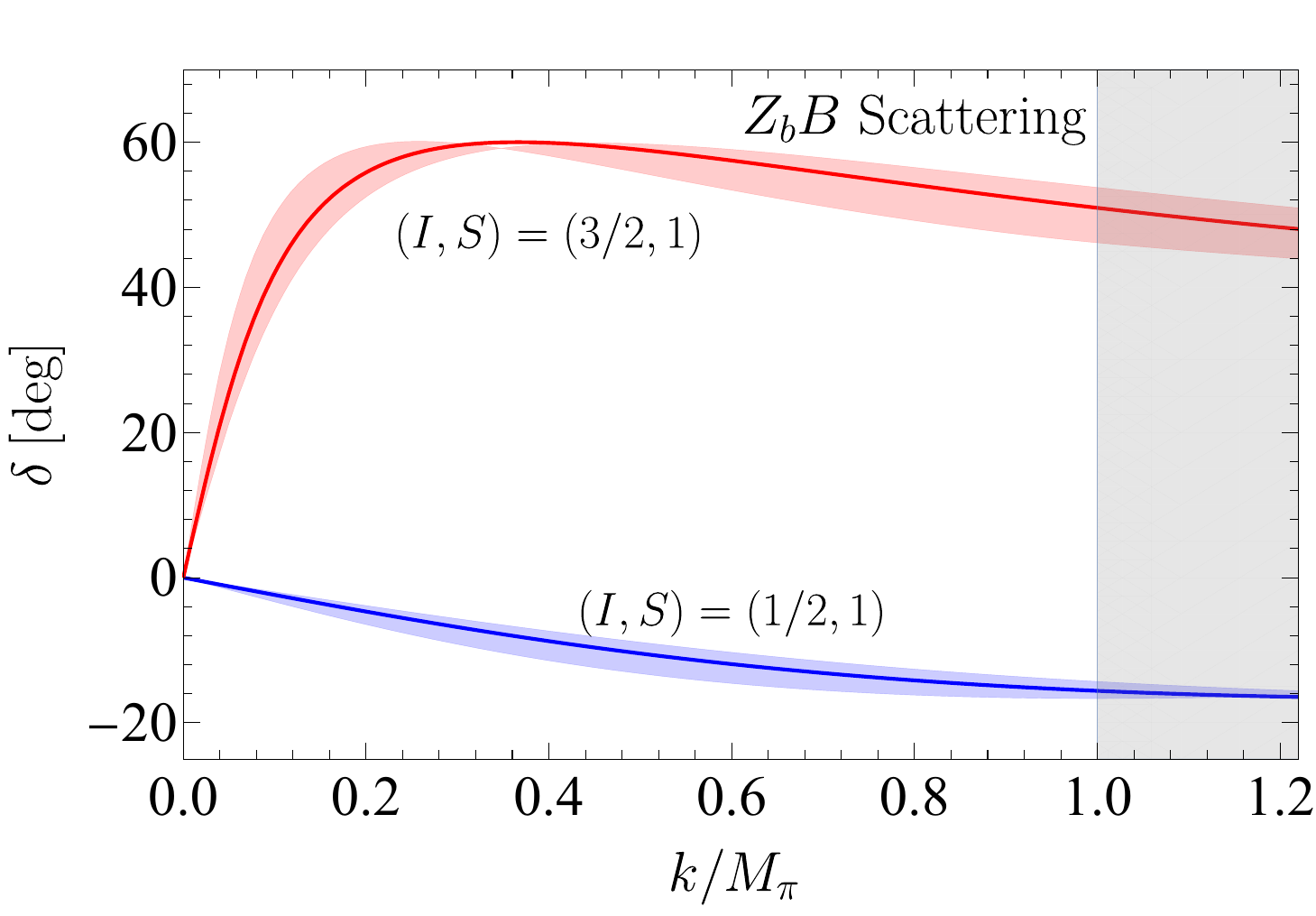}
\end{center}
\caption{\label{plot_Z_b-B_phaseshift_as_function_of_k}{$S$-wave phase shift $\delta$ as function of the momentum $k$ for the $I=3/2$ (solid line) and
    $I=1/2$ channels (dashed line) of elastic $Z_bB$ scattering.
    The bands are obtained by varying the input binding energies in the ranges given in
    Eq.~(\ref{eq:binding_energies}).
    Note that the pionless EFT expansion breaks down for momenta of order
    $M_\pi$ as indicated by the shaded area.
}}
\end{figure}
%
\subsection{Discussion of $\mathbf{Z_b B}$ scattering}
The elastic scattering $Z_bB$ is completely described by the formulae in Sec.~\ref{section_elastic_ZB_scattering} characterized by two observables: 
the $Z_bB$ scattering length $a_3$ and $S$-wave phase shift $\delta (k)$. The scattering length in the $I=3/2$, $S=1$ channel
is given by
\begin{align}
 a_3^{I=\frac{3}{2}, S=1} = -14.0^{+2.6}_{-6.0} \: \mbox{fm} \:,
\end{align}
and the corresponding phase shift in this channel is shown as a function of $k$ in Fig.~\ref{plot_Z_b-B_phaseshift_as_function_of_k}.
The bands are obtained by varying the input binding energies in the ranges given in Eq.~(\ref{eq:binding_energies}).
Note that the pionless EFT expansion breaks down for momenta of order
$M_\pi$ and our results in the shaded region of Fig.~\ref{plot_Z_b-B_phaseshift_as_function_of_k}
should only be taken as an indication of the general trend.
The positive phase shift corresponds indicates an attractive
interaction between the two scattered particles. However, as discussed above
the interaction is not strong enough to induce a three-body bound state.
For a increasing attraction of the meson-molecule interaction,
the scattering length $a_3$ tends to minus infinity and jumps to 
plus infinity when a bound state appears (see, e.g., Ref.~\cite{Hammer}).
The scattering length $a_3^{I=\frac{3}{2}, S=1}$ is large but negative
such that only a little more attraction would be needed to form a
universal trimer state.

From the negative phase shift in
Fig.~\ref{plot_Z_b-B_phaseshift_as_function_of_k}
for the $I=1/2$, $S=1$ channel, we conclude that the $ZB$ interaction in this
channel is weakly repulsive. The corresponding scattering length is
\begin{align}
 a_3^{I=\frac{1}{2}, S=1} = 0.6^{+0.3}_{-0.1} \: \mbox{fm} \:.
\end{align}
\subsection{Discussion of $\mathbf{Z_b B^*}$ scattering}
%
\begin{figure}[tb]
\begin{center}
    \includegraphics[width=0.48\textwidth]{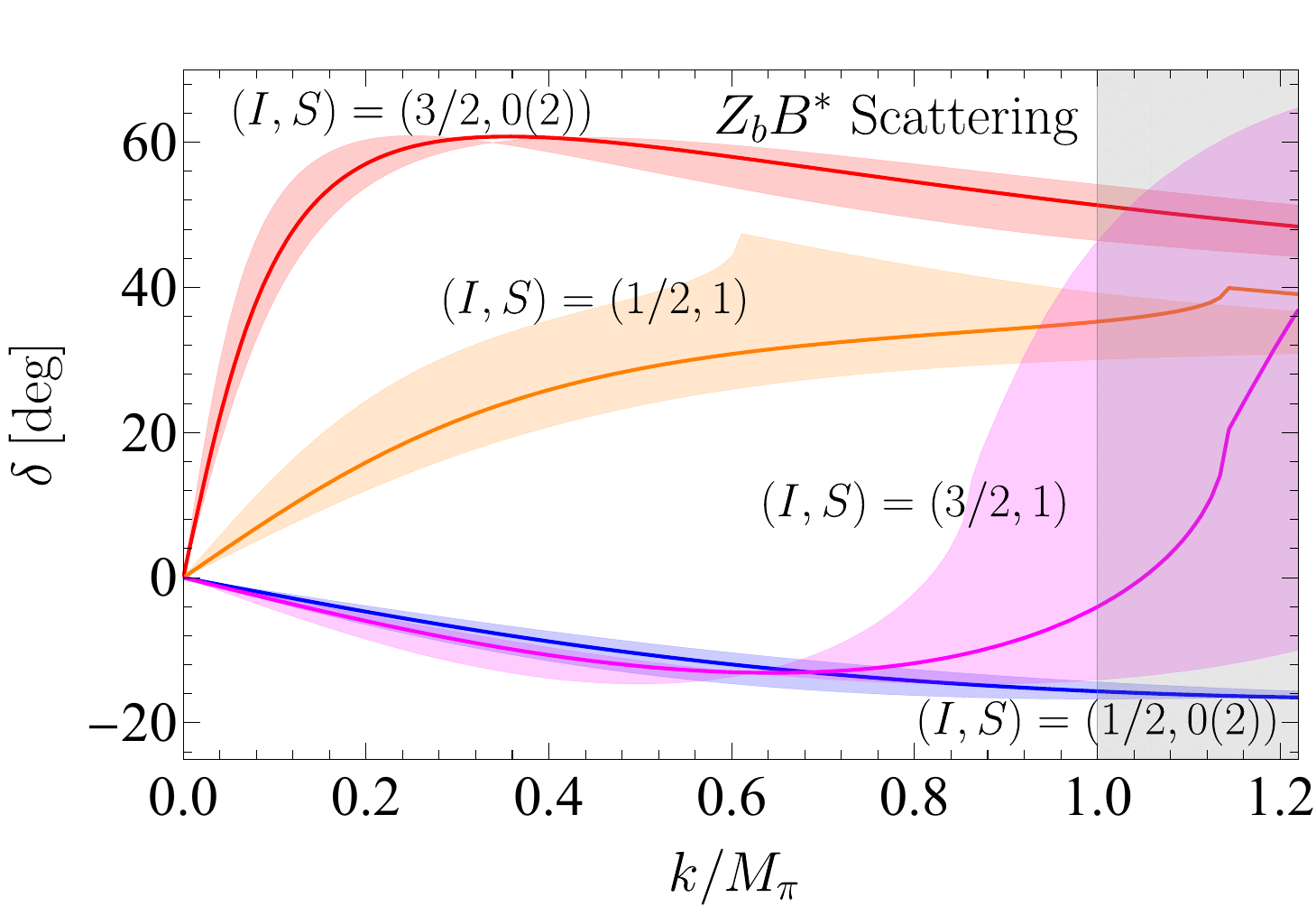}
\end{center}
\caption{\label{plot_Z_b-B_star_phaseshift_as_function_of_k}{$S$-wave phase shift $\delta$ as function of the momentum $k$ for all six channels in elastic
    $Z_bB^*$ scattering. The bands are obtained by varying the input binding energies in the ranges given in
    Eq.~(\ref{eq:binding_energies}). Note, that the $S = 0$ and $S = 2$ spin channels yield the same result. Moreover, the pionless EFT expansion breaks down for momenta of order
    $M_\pi$ as indicated by the shaded area.}}
\end{figure}
%
In the same way as for $Z_bB$ scattering,
one can analyze the scattering observables in the $Z_b B^*$ system.
We calculate the molecule-meson scattering length and the phase shift in
all six isospin-spin channels. Due the purely $S$-wave interaction at leading
order, the projection onto some of the
isospin and spin states leads to identical prefactors.
Therefore only four independent amplitudes remain.
The corresponding phase shifts are shown in
Fig.~\ref{plot_Z_b-B_star_phaseshift_as_function_of_k}.
Note that the pionless EFT expansion breaks down for momenta of order
$M_\pi$ indicated by the shaded region.
The $I=3/2,S=0,2$ and $I=1/2,S=1$ phase shifts indicate an attractive
interaction between the $Z_b$ and $B^*$. However, the attraction again is
not strong enough to produce trimer states. The
corresponding scattering lengths are:
\begin{align}
 a_3^{I=\frac{3}{2}, S=0,2} & = -14.8^{+2.8}_{-6.3} \: \mbox{fm} \:,\\
 a_3^{I=\frac{1}{2}, S=0,2} & = \phantom{-} 0.6^{+0.3}_{-0.1} \: \mbox{fm} \:,\\
 a_3^{I=\frac{3}{2}, S=1} & = \phantom{-}0.8^{+0.4}_{-0.1} \: \mbox{fm} \:,\\
 a_3^{I=\frac{1}{2}, S=1} & = -2.2^{+0.6}_{-1.5} \: \mbox{fm} \:,
\end{align}
in agreement with the absence of trimer states.
\subsection{Discussion of $\mathbf{Z'_b B^*}$ scattering}
%
\begin{figure}[tb]
\begin{center}
    \includegraphics[width=0.48\textwidth]{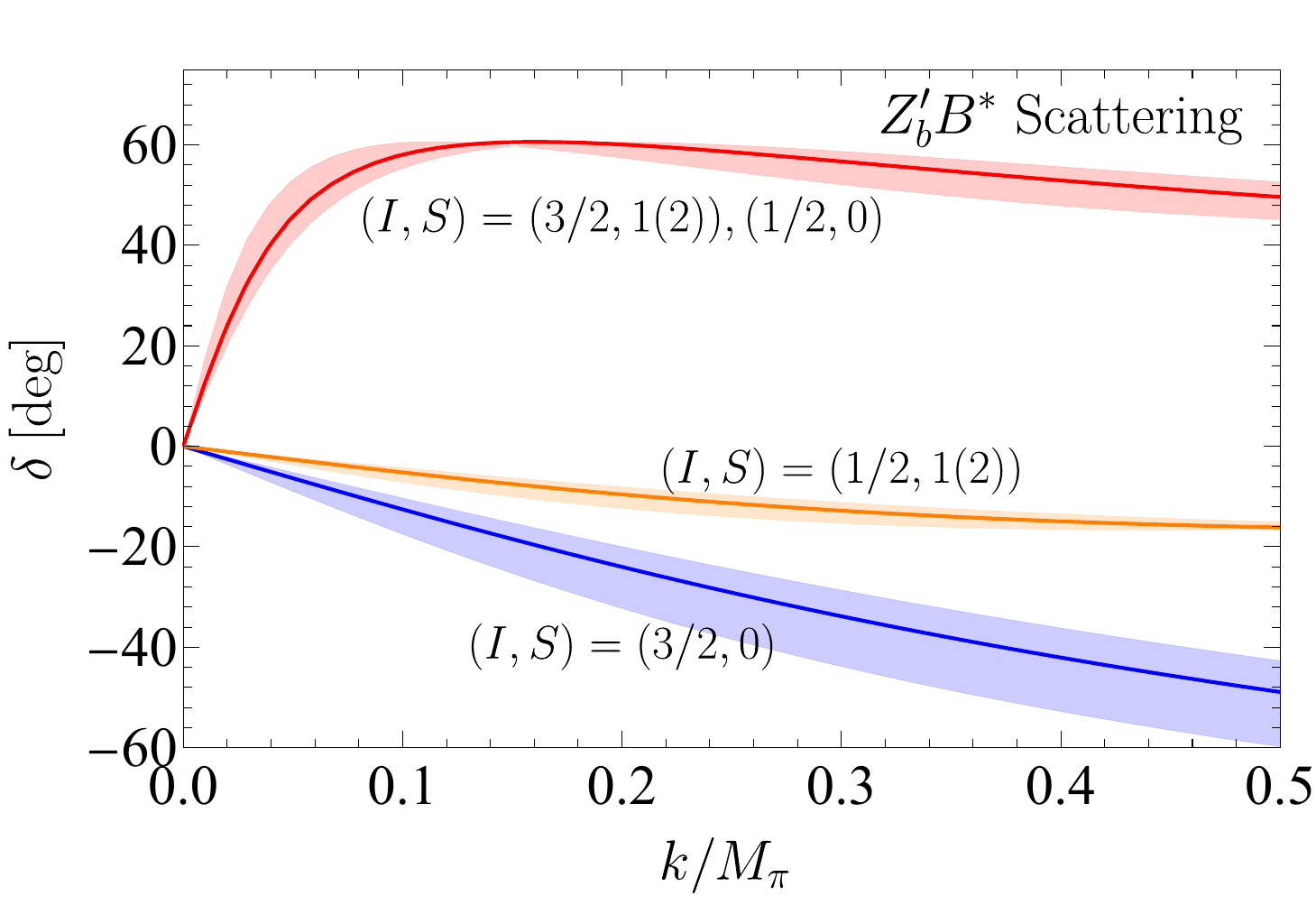}
\end{center}
\caption{\label{plot_Z_b_prime-B_star_phaseshift_as_function_of_k}{$S$-wave phase shift $\delta$ as function of the momentum $k$ for all six channels in elastic $Z'_bB^*$ scattering. The bands are obtained by varying the input binding energies in the ranges given in
    Eq.~(\ref{eq:binding_energies}). Note, that for each isospin state the $S = 1$ and $S = 2$ spin channels yield the same result and furthermore that the $I=1/2$, $S=0$ result is equivalent to that of $I=3/2$, $S=1,2$.}}
\end{figure}
%
Finally, we turn to $Z'_bB^*$ scattering.
In Fig.~\ref{plot_Z_b_prime-B_star_phaseshift_as_function_of_k}, we
show the $Z'_bB^*$ scattering phase shifts up to the $Z'_b$ breakup momentum,
$k\approx 0.5 M_\pi$ where the scattering is purely elastic. 
The meson-molecule scattering lengths in the different
spin-isospin channels are given by
\begin{align}
 a_3^{I=\frac{3}{2}, S=1,2} \: = \: a_3^{I=\frac{1}{2}, S=0} & = -32.5^{+6.1}_{-13.6} \: \mbox{fm} \:,\\
 a_3^{I=\frac{1}{2}, S=1,2} & = \phantom{-}1.3^{+0.6}_{-0.3} \: \mbox{fm} \:,\\
 a_3^{I=\frac{3}{2}, S=0} & = \phantom{-}3.2^{+1.4}_{-0.6} \: \mbox{fm} \:,
\end{align}
respectively. One observes that the absolute value of the scattering length in the $I=3/2$, $S=1,2$ and
$I=1/2$, $S=0$ channels is an order of magnitude larger than in all other processes and channels. The large negative value of
the scattering length reflects the steep rise of the phase shift
below $k\approx 0.1 M_\pi$. It  
indicates that the $I=3/2$, $S=1,2$ and 
$I=1/2$, $S=0$ channels in the $Z'_b B^*$ system are very close to
the emergence of trimer states due to the Efimov effect but
that the attraction is
not quite enough.

\section{Conclusions}
\label{section_conclusions}
In this work, we have investigated the bound states and
scattering processes of $B$ and $B^*$ mesons off the $Z_b (10610)$ and
the $Z'_b(10650)$. Using an pionless EFT with short-range contact interactions,
we have derived the integral equations for the corresponding scattering
amplitudes to leading order in the EFT expansion. Furthermore, we
investigated the ultraviolet behavior of the scattering amplitudes 
and ruled out the possibility of bound states of three bottom mesons due to the Efimov effect in all considered channels.
As a consequence, there are no three-body forces at leading order, and we were able to predict
the phase shifts and scattering lengths for the elastic scattering of $Z_b B$, $Z_b B^*$, 
and $Z'_b B^*$. Our analysis showed the
the $Z'_b B^*$ channel, in particular, is close to supporting an Efimov
state and has a very large scattering length. Our predictions could,
in principle, be tested via the final state interactions in the decays of
heavier particles into three $B/B^*$ mesons (cf.~the discussion in
Ref.~\cite{Canham_Hammer_3}) or in lattice simulations.
Because of the universality
of large scattering length physics, they apply to any system with
short-range interactions and the same spin-isospin structure.

In the future, it would be interesting to calculate the effective 
range corrections to our results. While this is straightforward in
principle, at present there is no experimental information on the effective
ranges available such that only order of magnitude estimates are
feasible. Since the $Z_b (10610)$ is at the border of applicability
of pionless EFT, an extension to include explicit pions analog to XEFT
for the $X(3872)$ in the charm sector~\cite{XEFT_1, XEFT_2} should be
considered.
With respect to future lattice calculations, it would then
also be interesting to
investigate the light quark mass dependence and finite volume effects
in a framework with explicit pions~\cite{Baru:2013rta,Jansen:2013cba,Jansen:2015lha}.
A related process is the short-distance production of three $B/B^*$ mesons. This process is also observable when the 
$Z_b$ and/or $Z_b'$ mesons are virtual states and will be 
considered in a forthcoming publication. For an investigation in the $D$ meson sector, see Ref.~\cite{Braaten:2021iot}.

Universal three-body states bound by the Efimov effect have been found
in various areas of physics, ranging from nuclear physics
to ultracold
atoms~\cite{Hammer:2010kp,Hammer:2017tjm,Naidon:2016dpf,Greene:2017cik}. 
While the search for hadronic molecules bound by
the Efimov effect has not been successful so far,
it remains an intriguing possibility
to form shallow three-body hadronic molecules with universal properties.

\section*{Acknowledgements}
H.-W.H. was supported by Deutsche Forschungsgemeinschaft (DFG, German Research Foundation) under Project ID 279384907 – SFB 1245 and by the German Federal Ministry of Education and Research (BMBF) (Grant No. 05P24RDB).
U.-G.M. was supported in part by the Chinese Academy of Sciences (CAS) 
President’s International Fellowship Initiative (PIFI) (Grant
No. 2025PD0022), by the MKW NRW under the funding code NW21-024-A, and  by the Deutsche Forschungsgemeinschaft (DFG,German Research Foundation) as part of the CRC 1639 NuMeriQS – project
no. 511713970. H.-L.F. is supported by the National Key R\&D Program of China under Grant No. 2023YFA1606703; by the National Natural Science Foundation of China (NSFC) under Grants No. 12125507, No. 12361141819, and No. 12047503.



\begin{thebibliography}{99}
  
\bibitem{Z_b_discovery_2}
  A.~Bondar {\it et al.} [Belle Collaboration],
  Phys.\ Rev.\ Lett.\  {\bf 108}, 122001 (2012)
  [arXiv:1110.2251 [hep-ex]].
  
\bibitem{Belle:2013urd}
P.~Krokovny \textit{et al.} [Belle],
Phys. Rev. D \textbf{88}, 052016 (2013)
[arXiv:1308.2646 [hep-ex]].

\bibitem{Belle:2014vzn}
A.~Garmash \textit{et al.} [Belle],
Phys. Rev. D \textbf{91}, 072003 (2015)
[arXiv:1403.0992 [hep-ex]].

\bibitem{ParticleDataGroup:2024cfk}
S.~Navas \textit{et al.} [Particle Data Group],
Phys. Rev. D \textbf{110}, 030001 (2024).

\bibitem{Belle:2015upu}
A.~Garmash \textit{et al.} [Belle],
Phys. Rev. Lett. \textbf{116}, 212001 (2016)
[arXiv:1512.07419 [hep-ex]].

\bibitem{Voloshin_2}
  A.~E.~Bondar, A.~Garmash, A.~I.~Milstein, R.~Mizuk and M.~B.~Voloshin,
  Phys.\ Rev.\ D {\bf 84}, 054010 (2011)
  [arXiv:1105.4473 [hep-ph]].

\bibitem{Sun}
  Z.~F.~Sun, J.~He, X.~Liu, Z.~G.~Luo and S.~L.~Zhu,
  Phys.\ Rev.\ D {\bf 84}, 054002 (2011)
  [arXiv:1106.2968 [hep-ph]].


  
\bibitem{Yang}
  Y.~Yang, J.~Ping, C.~Deng and H.~S.~Zong,
  J.\ Phys.\ G {\bf 39}, 105001 (2012)
  [arXiv:1105.5935 [hep-ph]].
  
\bibitem{Zhang}
  J.~R.~Zhang, M.~Zhong and M.~Q.~Huang,
  Phys.\ Lett.\ B {\bf 704}, 312 (2011)
  [arXiv:1105.5472 [hep-ph]].

\bibitem{Cui}
C.~Y.~Cui, Y.~L.~Liu and M.~Q.~Huang,
Phys.\ Rev.\ D {\bf 85}, 074014 (2012)
[arXiv:1107.1343 [hep-ph]].
  
\bibitem{Cleven}
  M.~Cleven, F.~K.~Guo, C.~Hanhart and U.-G.~Mei\ss ner,
  Eur.\ Phys.\ J.\ A {\bf 47}, 120 (2011)
  [arXiv:1107.0254 [hep-ph]].

\bibitem{Cleven_2}
  M.~Cleven, Q.~Wang, F.~K.~Guo, C.~Hanhart, U.-G.~Mei\ss ner and Q.~Zhao,
  Phys.\ Rev.\ D {\bf 87}, 074006 (2013)
  [arXiv:1301.6461 [hep-ph]].

\bibitem{Zhao:2021cvg}
M.~J.~Zhao, Z.~Y.~Wang, C.~Wang and X.~H.~Guo,
Phys. Rev. D \textbf{105}, 096016 (2022)
[arXiv:2112.12633 [hep-ph]].

\bibitem{He:2024aej}
W.~He, D.~S.~Zhang and Z.~F.~Sun,
Phys. Rev. D \textbf{110} (2024), 054006 (2024)
[arXiv:2403.02099 [hep-ph]].

\bibitem{Yalikun:2025ssz}
N.~Yalikun, X.~K.~Dong and U.-G.~Mei\ss{}ner,
[arXiv:2503.01322 [hep-ph]].

\bibitem{Ali}
  A.~Ali, C.~Hambrock and W.~Wang,
  Phys.\ Rev.\ D {\bf 85}, 054011 (2012)
  [arXiv:1110.1333 [hep-ph]].

\bibitem{Guo}
  T.~Guo, L.~Cao, M.~Z.~Zhou and H.~Chen,
  arXiv:1106.2284 [hep-ph].

\bibitem{Bugg}
  D.~V.~Bugg,
  Europhys.\ Lett.\  {\bf 96}, 11002 (2011)
  [arXiv:1105.5492 [hep-ph]].

  
\bibitem{Chen-Liu}
  D.~Y.~Chen and X.~Liu,
  Phys.\ Rev.\ D {\bf 84}, 094003 (2011)
  [arXiv:1106.3798 [hep-ph]].
  
\bibitem{Sadl:2021bme}
M.~Sadl and S.~Prelovsek,
Phys. Rev. D \textbf{104}, 114503 (2021)
[arXiv:2109.08560 [hep-lat]].

\bibitem{Guo:2016bjq} 
  F.-K.~Guo, C.~Hanhart, Y.~S.~Kalashnikova, P.~Matuschek, R.~V.~Mizuk, A.~V.~Nefediev, Q.~Wang and J.-L.~Wynen,
  Phys.\ Rev.\ D {\bf 93}, 074031 (2016)
  [arXiv:1602.00940 [hep-ph]].

\bibitem{Hanhart:2015cua} 
  C.~Hanhart, Y.~S.~Kalashnikova, P.~Matuschek, R.~V.~Mizuk, A.~V.~Nefediev and Q.~Wang,
  Phys.\ Rev.\ Lett.\  {\bf 115}, 202001 (2015)
  [arXiv:1507.00382 [hep-ph]].
  
\bibitem{Wang:2018jlv}
Q.~Wang, V.~Baru, A.~A.~Filin, C.~Hanhart, A.~V.~Nefediev and J.~L.~Wynen,
Phys. Rev. D \textbf{98}, 074023 (2018)
[arXiv:1805.07453 [hep-ph]].

\bibitem{Baru:2019xnh}
V.~Baru, E.~Epelbaum, A.~A.~Filin, C.~Hanhart, A.~V.~Nefediev and Q.~Wang,
Phys. Rev. D \textbf{99}, 094013 (2019)
[arXiv:1901.10319 [hep-ph]].

\bibitem{Krug:2020ufl}
S.~L.~Krug and C.~Hanhart,
Bull. Lebedev Phys. Inst. \textbf{47}, 334 (2020).

\bibitem{Baru:2020ywb}
V.~Baru, E.~Epelbaum, A.~A.~Filin, C.~Hanhart, R.~V.~Mizuk, A.~V.~Nefediev and S.~Ropertz,
Phys. Rev. D \textbf{103}, 034016 (2021)
[arXiv:2012.05034 [hep-ph]].

\bibitem{Chen:2015jgl} 
  Y.~H.~Chen, J.~T.~Daub, F.~K.~Guo, B.~Kubis, U.-G.~Mei{\ss}ner and B.~S.~Zou,
  Phys.\ Rev.\ D {\bf 93}, 034030 (2016)
  [arXiv:1512.03583 [hep-ph]].

\bibitem{Chen:2016mjn} 
  Y.~H.~Chen, M.~Cleven, J.~T.~Daub, F.~K.~Guo, C.~Hanhart, B.~Kubis, U.-G.~Mei\ss ner and B.~S.~Zou,
  Phys.\ Rev.\ D {\bf 95}, 034022 (2017)
  [arXiv:1611.00913 [hep-ph]].

 \bibitem{Guo:2017jvc}
 F.~K.~Guo, C.~Hanhart, U.-G.~Mei\ss{}ner, Q.~Wang, Q.~Zhao and B.~S.~Zou,
 Rev. Mod. Phys. \textbf{90}, 015004 (2018)
 [erratum: Rev. Mod. Phys. \textbf{94}, 029901 (2022)]
 [arXiv:1705.00141 [hep-ph]].

  
\bibitem{Braaten_1}
  E.~Braaten, C.~Langmack and D.~H.~Smith,
  Phys.\ Rev.\ D {\bf 90}, 014044 (2014)
  [arXiv:1402.0438 [hep-ph]].
  
  \bibitem{Prelovsek:2019ywc}
  S.~Prelovsek, H.~Bahtiyar and J.~Petkovic,
  Phys. Lett. B \textbf{805}, 135467 (2020)
  [arXiv:1912.02656 [hep-lat]].
  
\bibitem{Hoffmann:2024hbz}
J.~Hoffmann and M.~Wagner,
[arXiv:2412.06607 [hep-lat]].

  
\bibitem{Guo:2010ak}
  F.~K.~Guo, C.~Hanhart, G.~Li, U.-G.~Mei\ss ner and Q.~Zhao,
  Phys.\ Rev.\ D {\bf 83}, 034013 (2011)
  [arXiv:1008.3632 [hep-ph]].
 
  \bibitem{Canham_Hammer_3}
  D.~L.~Canham, H.-W.~Hammer and R.~P.~Springer,
  Phys.\ Rev.\ D {\bf 80}, 014009 (2009)
  [arXiv:0906.1263 [hep-ph]].

\bibitem{Braaten_Kusunoki}
  E.~Braaten and M.~Kusunoki,
  Phys.\ Rev.\ D {\bf 69}, 074005 (2004)
  [hep-ph/0311147].


\bibitem{Wilbring_Hammer_Meissner}
  E.~Wilbring, H.-W.~Hammer and U.-G.~Mei\ss ner,
  Phys.\ Lett.\ B {\bf 726}, 326 (2013)
  [arXiv:1304.2882 [hep-ph]].

\bibitem{vanKolck:1997ut} 
  U.~van Kolck,
  Lect.\ Notes Phys.\  {\bf 513}, 62 (1998)
  [hep-ph/9711222].

  
\bibitem{van_Kolck_2}
  U.~van Kolck,
  Nucl.\ Phys.\ A {\bf 645}, 273 (1999)
  [nucl-th/9808007].

  
\bibitem{Kaplan_3}
  D.~B.~Kaplan, M.~J.~Savage and M.~B.~Wise,
  Phys.\ Lett.\ B {\bf 424}, 390 (1998)
  [nucl-th/9801034].

\bibitem{Kaplan_4}
  D.~B.~Kaplan, M.~J.~Savage and M.~B.~Wise,
  Nucl.\ Phys.\ B {\bf 534}, 329 (1998)
  [nucl-th/9802075].


\bibitem{Bedaque:1998kg} 
  P.~F.~Bedaque, H.-W.~Hammer and U.~van Kolck,
  Phys.\ Rev.\ Lett.\  {\bf 82}, 463 (1999)
  [nucl-th/9809025].
  
\bibitem{Bedaque:1998km} 
  P.~F.~Bedaque, H.-W.~Hammer and U.~van Kolck,
  Nucl.\ Phys.\ A {\bf 646}, 444 (1999)
  [nucl-th/9811046].

\bibitem{Bedaque:1999ve} 
  P.~F.~Bedaque, H.-W.~Hammer and U.~van Kolck,
  Nucl.\ Phys.\ A {\bf 676}, 357 (2000)
  [nucl-th/9906032].

\bibitem{Efimov}
  V.~Efimov,
  Phys.\ Lett.\ B {\bf 33}, 563 (1970).

\bibitem{Bedaque:1997qi} 
  P.~F.~Bedaque and U.~van Kolck,
  Phys.\ Lett.\ B {\bf 428}, 221 (1998)
  [nucl-th/9710073].

\bibitem{Bedaque:1998mb} 
  P.~F.~Bedaque, H.-W.~Hammer and U.~van Kolck,
  Phys.\ Rev.\ C {\bf 58}, R641 (1998)
  [nucl-th/9802057].

\bibitem{Epelbaum:2016ffd} 
  E.~Epelbaum, J.~Gegelia, U.-G.~Mei{\ss}ner and D.~L.~Yao,
  Eur.\ Phys.\ J.\ A {\bf 53}, 98 (2017)
  [arXiv:1611.06040 [nucl-th]].
  
\bibitem{Griesshammer:2005ga} 
  H.~W.~Griesshammer,
  Nucl.\ Phys.\ A {\bf 760}, 110 (2005)
  [nucl-th/0502039].
  
\bibitem{Hammer}
  E.~Braaten and H.-W.~Hammer,
  Phys.\ Rept.\  {\bf 428}, 259 (2006)
  [cond-mat/0410417].

\bibitem{Hammer:2010kp} 
  H.-W.~Hammer and L.~Platter,
  Ann.\ Rev.\ Nucl.\ Part.\ Sci.\  {\bf 60}, 207 (2010)
  [arXiv:1001.1981 [nucl-th]].

\bibitem{Garcilazo:2018rwu}
H.~Garcilazo and A.~Valcarce,
Phys. Lett. B \textbf{784}, 169 (2018)
[arXiv:1808.00226 [hep-ph]].

\bibitem{Ma:2018vhp}
L.~Ma, Q.~Wang and U.-G.~Mei\ss{}ner,
Phys. Rev. D \textbf{100}, 014028 (2019)
[arXiv:1812.09750 [hep-ph]].

\bibitem{Deng:2024pwm}
C.~R.~Deng and C.~S.~An,
Phys. Rev. D \textbf{111}, 034002 (2025)
[arXiv:2411.03589 [hep-ph]].
  
\bibitem{Hammer:2000nf} 
  H.-W.~Hammer and T.~Mehen,
  Nucl.\ Phys.\ A {\bf 690}, 535 (2001)
  [nucl-th/0011024].

\bibitem{Kaspschak:2021hbc}
B.~Kaspschak and U.-G.~Mei\ss{}ner,
Mach. Learn. Sci. Tech. \textbf{3}, 025003 (2022)
[arXiv:2111.07820 [nucl-th]].

\bibitem{Wilbring-Diss}E.~Wilbring, Efimov Effect in Pionless Effective Field Theory and its Application to Hadronic Molecules, PhD thesis, University of Bonn (2016).

\bibitem{Baru:2017gwo}
V.~Baru, E.~Epelbaum, A.~A.~Filin, C.~Hanhart and A.~V.~Nefediev,
JHEP \textbf{06}, 158 (2017)
[arXiv:1704.07332 [hep-ph]].

\bibitem{XEFT_1}
  S.~Fleming, M.~Kusunoki, T.~Mehen and U.~van Kolck,
  Phys.\ Rev.\ D {\bf 76}, 034006 (2007)
  [hep-ph/0703168].


  
\bibitem{XEFT_2}
  E.~Braaten, H.-W.~Hammer and T.~Mehen,
  Phys.\ Rev.\ D {\bf 82}, 034018 (2010)
  [arXiv:1005.1688 [hep-ph]].

\bibitem{Baru:2013rta} 
  V.~Baru, E.~Epelbaum, A.~A.~Filin, C.~Hanhart, U.-G.~Mei\ss ner and A.~V.~Nefediev,
  Phys.\ Lett.\ B {\bf 726}, 537 (2013)
  [arXiv:1306.4108 [hep-ph]].

  
\bibitem{Jansen:2013cba} 
  M.~Jansen, H.-W.~Hammer and Y.~Jia,
  Phys.\ Rev.\ D {\bf 89}, 014033 (2014)
  [arXiv:1310.6937 [hep-ph]].

\bibitem{Jansen:2015lha} 
  M.~Jansen, H.-W.~Hammer and Y.~Jia,
  Phys.\ Rev.\ D {\bf 92}, 114031 (2015)
  [arXiv:1505.04099 [hep-ph]].

\bibitem{Braaten:2021iot}
E.~Braaten and H.~W.~Hammer,
Phys. Rev. Lett. \textbf{128}, 032002 (2022)
[arXiv:2107.02831 [hep-ph]].

\bibitem{Hammer:2017tjm}
H.~W.~Hammer, C.~Ji and D.~R.~Phillips,
J. Phys. G \textbf{44}, 103002 (2017)
[arXiv:1702.08605 [nucl-th]].

\bibitem{Naidon:2016dpf}
P.~Naidon and S.~Endo,
Rept. Prog. Phys. \textbf{80}, 056001 (2027)
[arXiv:1610.09805 [quant-ph]].

  
\bibitem{Greene:2017cik}
C.~H.~Greene, P.~Giannakeas and J.~Perez-Rios,
Rev. Mod. Phys. \textbf{89}, 035006 (2017)
[arXiv:1704.02029 [cond-mat.quant-gas]].



\end{thebibliography}
\end{document}